%% file: main.tex
\PassOptionsToPackage{dvipsnames}{xcolor}
\documentclass{article}
\usepackage[preprint]{icml2026}
\input{macros}

\def\mytitle{Speculative Decoding for Semantic-ID-Based Generative Recommendations}
\icmltitlerunning{\mytitle}

\begin{document}

\twocolumn[
\icmltitle{\mytitle}
\begin{icmlauthorlist}
\icmlauthor{Harshal A. Chaudhari}{etsy}
\end{icmlauthorlist}
\icmlaffiliation{etsy}{Etsy Inc., Brooklyn, NY, USA}
\icmlcorrespondingauthor{Harshal A. Chaudhari}{hchaudhari@etsy.com}
\icmlkeywords{generative recommendation, semantic IDs, constrained beam search, speculative decoding, tree attention}
\vskip 0.3in
]

\printAffiliationsAndNotice{}

\input{sections/00_abstract}
\input{sections/00_highlight_figure}
\input{sections/01_introduction}
\input{sections/02_decoder_and_certificate}
\input{sections/03_catalog_trie_structure}
\input{sections/04_decoding_under_budget}
\input{sections/05_experiments}
\input{sections/06_related_work}
\input{sections/07_impact_statement}

\setlength{\bibsep}{3pt}
\bibliographystyle{icml2026}
\bibliography{references}

\newpage
\appendix
\crefalias{section}{appendix}
\onecolumn
\input{sections/appendix_A_assumptions_and_proofs}
\input{sections/appendix_B_further_guarantees}
\input{sections/appendix_C_computing_announcement}
\input{sections/appendix_D_details_of_experiments}
\input{sections/appendix_E_faq}
\end{document}

%% file: macros.tex
\usepackage[T1]{fontenc}
\usepackage{amsthm,amssymb,amsmath,mathtools,bm}
\usepackage{xcolor}

\usepackage{graphicx}
\usepackage{booktabs,multirow,subcaption,enumitem}
\usepackage{pifont}
\usepackage{xspace}            
\usepackage{tikz}
\usetikzlibrary{arrows.meta,positioning,calc}
\usepackage{algorithm}
\usepackage{algorithmic}
\usepackage[font={small,stretch=1.05},labelfont=bf]{caption}
\usepackage[USenglish]{babel}
\usepackage[final]{microtype}
\usepackage[colorlinks=true]{hyperref}
\hypersetup{linkcolor=Cerulean,citecolor=Cerulean,urlcolor=Cerulean,filecolor=Cerulean}
\usepackage[capitalize,nameinlink]{cleveref}
\graphicspath{{./figures/}}

\setlist{itemsep=1pt,topsep=2pt,parsep=0pt,leftmargin=1.5em}

\theoremstyle{plain}
\newtheorem{theorem}{Theorem}
\newtheorem{proposition}{Proposition}
\newtheorem{lemma}{Lemma}
\theoremstyle{definition}
\newtheorem{remark}{Remark}
\crefname{remark}{Remark}{Remarks}
\crefname{proposition}{Proposition}{Propositions}

\newcommand{\beq}[1]{\begin{equation}#1\end{equation}}

\DeclarePairedDelimiter\floor{\lfloor}{\rfloor}
\DeclarePairedDelimiter\ceil{\lceil}{\rceil}

\newcommand{\method}{SIDECAR}
\newcommand{\bmethod}{BUDGET-\method{}}
\newcommand{\cE}{\mathcal{E}}
\newcommand{\cT}{\mathcal{T}}

\newcommand{\sid}{\mathrm{SID}}
\newcommand{\children}{\mathrm{children}}
\newcommand{\depth}{\mathrm{depth}}

\newcommand{\ModelK}{\ensuremath{\text{\textsc{Model-K}}}\xspace}
\newcommand{\BeamK}{\ensuremath{\text{\textsc{Beam-K}}}\xspace}
\newcommand{\SidecarK}{\ensuremath{\text{\textsc{Sidecar-K}}}\xspace}
\newcommand{\SidecarKr}[1]{\ensuremath{\text{\textsc{Sidecar-K}}_{#1}}\xspace}
\newcommand{\Exact}{\textsc{Exact}\xspace}
\newcommand{\Certificate}{\textsc{Certificate}\xspace}

\newcommand{\gap}{\mathrm{gap}}

\newcommand{\cmark}{\textcolor{green!55!black}{\ding{51}}}
\newcommand{\xmark}{\textcolor{red!75!black}{\ding{55}}}
\newcommand{\smark}{\textcolor{gray}{\raisebox{-0.1ex}{$\sim$}}}
\newcommand{\namark}{\textcolor{gray}{\scriptsize N/A}}
\newcommand{\hdr}[1]{\begin{tabular}[b]{@{}c@{}}#1\end{tabular}}

\makeatletter
\newcommand{\algline}[1]{\protected@edef\@currentlabel{\arabic{ALC@line}}\label{#1}}
\makeatother

\tikzset{ pfnd/.style={draw, rounded corners=2pt, minimum width=6mm, minimum height=4.6mm, inner sep=1pt,
	font=\scriptsize}, pfopen/.style={pfnd, fill=blue!20}, pfclosed/.style={pfnd, fill=white, draw=black!45},
	pffed/.style={pfnd, fill=blue!20, thick}, pfitem/.style={draw, minimum width=3.4mm, minimum height=3.4mm, inner
	sep=0pt, fill=gray!15}, pfitemc/.style={pfitem, thick, fill=white}, pfslot/.style={draw, minimum width=4.4mm,
	minimum height=4.4mm, inner sep=0pt}, pffull/.style={pfslot, fill=blue!30}, pfempty/.style={pfslot, fill=white,
	draw=black!40}, pflvl/.style={font=\scriptsize\itshape, text=black!60, anchor=east}, pfann/.style={font=\scriptsize,
	text=black!70, align=left, anchor=west}, pfttl/.style={font=\footnotesize\bfseries, align=center}}

%% file: sections/00_abstract.tex
\begin{abstract}
Generative recommender systems with Semantic IDs (SIDs) represent each item as a short sequence of $L$ discrete tokens
	and use a sequence model to predict the next item's tokens from a user's history.
Industry-scale deployments typically use these models to retrieve multiple candidates per request with constrained beam
	search.
However, this incurs a fixed cost of $L$ sequential model calls per request, and accuracy degrades as the SID generator
	becomes less sharp, without a guarantee on how close the output is to the exact top-$K$.
Production systems therefore cannot detect degraded retrieval for individual requests.

We propose \method{}, which uses approximate draft candidates and a tree-attention pass to score the entire draft in a
	single model call, memory permitting.
This pass bounds the probability of every unexplored SID prefix and certifies the exact top-$K$ if the draft contains
	them.
Otherwise, a branch-and-bound fallback expands only the parts of the trie that can still change the output, reaching
	certification within at most $L$ calls for any draft.
Moreover, under a budget of $M$ tokens per call and $Y$ calls, which for $Y = L$ matches the resources of a beam search
	of width $M$, \method{} announces a factor $1 + \delta$ within which its list will be exact at every rank.
Depending on model sharpness, \method{} returns the exact top-$K$ using a fraction of beam search's calls or memory.
Even when the budget is too small to recover the exact top-$K$, \method{}, unlike beam search, reports the announced
	factor, allowing the serving system to stop early once a quality target or latency limit is reached.
\end{abstract}

%% file: sections/00_highlight_figure.tex
\begin{figure}[t]
\centering
\scriptsize
\setlength{\tabcolsep}{3pt}
\renewcommand{\arraystretch}{1.12}
\resizebox{\columnwidth}{!}{%
\begin{tabular}{@{}l c c c c@{}}
\toprule
 & \hdr{Exhaustive\\scoring} & \hdr{Beam-search\\decoders} & \hdr{Speculative\\decoders} & \hdr{\method{}\\(ours)} \\
\midrule
Exact top-$K$ & \cmark & \smark & \smark & \cmark \\
Exactness certificate & \cmark & \xmark & \xmark & \cmark \\
Early stop & \namark & \xmark & \xmark & \cmark \\
Guarantee under a budget & \namark & \xmark & \xmark & \cmark \\
Sequential model calls & 1 & $L$ & $< L$ & $\sim$1--2 (worst case $L$) \\
Memory per call & catalog & $W$ & $W$ + draft & $M$ ($\le W$) \\
Draft & --- & --- & trained or retrieved draft & any draft candidates \\
\midrule
Examples & --- & \hdr{TIGER, GR4AD,\\UtilityGR} & \hdr{AtSpeed, NEZHA,\\SpecGR} & --- \\
\bottomrule
\end{tabular}}
\vspace{2pt}
\includegraphics[width=\columnwidth]{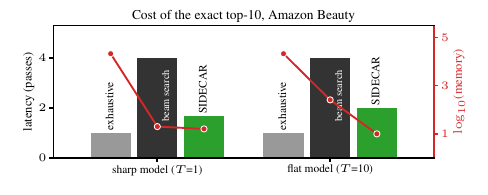}
\caption{\method{} returns the top-$K$ with an exactness certificate, in fewer passes and less memory per pass than
	beam-search decoders.}
\label{fig:highlight}
\end{figure}

%% file: sections/01_introduction.tex
\section{Introduction}
\label{sec:intro}
\input{sections/fig_example}

Semantic IDs (SIDs) have become a common way for large-scale recommender systems to represent their items: a residual
	quantizer maps the content of each item to $L$ discrete codes~\citep{lee2022rqvae,rajput2023tiger}, and a sequence
	model emits these codes one after another, as generative retrieval does for
	documents~\citep{tay2022dsi,wang2022nci}.
Thus, the model does not need an output layer of the size of the catalog, and a new item receives an identifier from its
	content alone.
Following TIGER, most research has focused on generating these codes, learning them with collaborative
	signals~\citep{jin2024lmindexer,wang2024letter,liu2025etegrec}, and several companies now serve this design in
	production~\citep{he2025plum,damico2026glide,deng2025onerec,zhai2024actions}.

Decoding has also received attention; however, most of this work changes the model to suit the decoder, with multi-token
	heads~\citep{hou2025rpg,zhan2026stamp}, trained drafters or self-drafting
	heads~\citep{lin2025atspeed,wang2026nezha}, or scorers retrained for beam search~\citep{zhuo2020beamtree}.
A model directly served after training is still decoded with constrained beam search over the trie of valid
	SIDs~\citep{ju2025handbook}, and its large-scale variants adapt the width~\citep{gr4ad}, split the
	beam~\citep{multidecoderonerec}, or rescore prefixes~\citep{utilitygr}, still running one dependent pass per SID
	level.
This comes at a measurable cost: a single head decoded with beam search produces better rankings and content quality,
	whereas multi-token prediction provides higher throughput~\citep{oh2026personalized}.
We therefore ask the following question: can we design a decoder that returns the exact output of a single head while
	avoiding its sequential cost?

Beam search spends memory and latency, the two constrained resources of large-scale production systems, in a rigid way:
	on every request it runs $L$ passes of $W$ prefixes each and outputs a list only after the last pass.
Yet most of these passes do not change the final output.
The SID codebooks address far more cells than most catalog sizes, so below the first levels most prefixes contain single
	items.
Beam search nevertheless spends a full sequential pass on each of the remaining codes of such prefixes.
This approach also constrains the design of the codes themselves.
Since every level adds a dependent pass, SIDs are usually kept short~\citep{hou2025rpg,xie2026singlelevel}, and recent
	industrial work moves capacity from depth to a larger codebook~\citep{xie2026singlelevel}.
A larger codebook, however, makes the code space dwarf the catalog even more.
Whether deeper trees would capture the embedding space better remains open.
Moreover, beam search cannot tell which of its output lists differ from the exact top-$K$ of the model.
Its output list is exact when $W$ exceeds the number of prefixes above the $K$-th best item at every level, which beam
	search cannot verify at runtime.
In \cref{fig:example}(b), for example, a beam of width $K$ prunes the prefix that holds the second-best item, and
	nothing in the computation of beam search reveals that its list is therefore wrong.
Both limitations above, namely the fixed cost of $L$ passes and the missing certificate, stem from the decoder rather
	than from the model itself.
Hence, we keep the model and its codes unchanged and redesign only the decoder.

By design, beam search assumes that nothing is known about the final output before its first pass.
However, most large-scale recommender systems in the industry can easily provide a draft of the answer.
Most contemporary production stacks run a first-stage retriever before the ranker, and candidates are often computed
	offline using batched jobs.
Alternatively, a batch job can store the top-$N$ items of the model itself, or of a cheaper, heavily quantized copy of
	the model that is exported at training time.
The batch job can compute a high-quality draft for each user from an interaction history that is a few interactions old.
At request time, the decoder can then retrieve such drafts through KV stores.
Furthermore, one pass with a tree-attention mask~\citep{miao2024specinfer,cai2024medusa} can score the whole draft, just
	as speculative decoders verify a draft tree~\citep{leviathan2023speculative}.
In this paper, we introduce \method{} (Semantic-ID Decoding with Exactness Certificates from Approximate Retrieval
	drafts), a decoder that exploits such drafts.
Because the decoding is constrained to the trie of valid SIDs~\citep{decao2021genre}, the pass that scores the draft
	also bounds every item outside it.
Hence, \method{} can prove that its list is the exact top-$K$ (\cref{fig:example}(a)), which no existing speculative
	decoder for SIDs does~\citep{lin2025atspeed,ding2026specgr}, and when the certification fails, a fallback expands
	only the prefixes that can further improve the output list.
Thus, given enough memory per pass, the output is exact for every draft.
The draft quality itself decides only how many passes are required.
Moreover, \method{} uses the structure of the catalog trie, which beam search ignores, to announce how close to exact
	its output will be as early as after the first pass.
Finally, since \method{} does not pay one pass per SID level on every request, we argue that it opens new avenues for
	deeper SID trees, which can capture richer semantic structures from the item embeddings.

We develop the theory of \method{} in two stages.
In \cref{sec:onepass,sec:trie}, we assume unbounded memory per pass, where the certificate, the fallback, and the role
	of the catalog trie are simplest to state.
Building upon this intuition, \cref{sec:budgets} adds the memory and latency budgets of serving, under which \bmethod{}
	plans its passes so that its list comes with a worst-case guarantee within the budget.
Our experiments (\cref{sec:exp}) compare the two decoders at beam search's width and its $L$ passes, across models of
	increasing flatness.
On a sharp model, the decoders return the same lists, but \method{} needs less than half of the passes at equal memory
	and certifies every exact list, whereas beam search certifies none.
As the model flattens, beam search's correctness collapses at every width short of the whole catalog, because it ranks
	prefixes by their probability mass, whereas the top items of a flat model lie in clusters with little mass.
\method{}, in contrast, keeps returning the exact list with a fraction of beam search's memory, since its draft points
	directly to the top items, whereas beam search has to keep every heavy cluster to reach them.
From half of beam search's width on, \method{} also certifies that its list is exact.
Finally, the announcement after the first call tells the deployment which of these cases it is in, and it states the
	guarantee factor where neither resource suffices, whereas beam search gives no indication of how far its list is
	from the exact one.
\Cref{fig:highlight} compares \method{} with exhaustive scoring, beam-search decoders, and speculative decoders.
Appendix~\ref{app:proofs} provides the assumptions and the proofs of all formal results.

%% file: sections/fig_example.tex
\begin{figure*}[t]
\centering
\resizebox{\textwidth}{!}{%
\begin{tikzpicture}[
  font=\small, every node/.style={inner sep=2pt},
  nd/.style={draw, rounded corners=2pt, minimum width=9mm, minimum height=5.5mm, align=center},
  root/.style={nd, fill=gray!20}, fed/.style={nd, fill=blue!15}, cand/.style={nd, thick, fill=white},
  sib/.style={nd, fill=green!15}, front/.style={nd, dashed, fill=red!12},
  kept/.style={nd, fill=blue!15}, pruned/.style={nd, dashed, draw=black!45, text=black!45, fill=white},
  scored/.style={nd, fill=white}, ret/.style={nd, thick, fill=white},
  miss/.style={nd, dashed, draw=red!70, text=red!70, fill=white},
  ed/.style={font=\scriptsize, fill=white, inner sep=1pt}, edp/.style={ed, text=black!45},
  pr/.style={font=\scriptsize}, prp/.style={pr, text=black!45}, prm/.style={pr, text=red!70},
  lvl/.style={font=\scriptsize\itshape, text=black!60}, ttl/.style={font=\small\bfseries}]
\node[ttl] at (4.3,0.9) {(a) \method{}: one pass over the draft $\{57, 54, 14\}$};
\node[root] (root) at (4.3,0) {root ($h$)};
\node[fed] (n12) at (1.95,-1.4) {12};
\node[fed] (n13) at (6.675,-1.4) {13};
\node[fed] (n201) at (0.65,-2.8) {201};
\node[fed] (n200) at (3.25,-2.8) {200};
\node[fed] (n123) at (5.85,-2.8) {123};
\node[front] (n120) at (7.5,-2.8) {120};
\node[cand, label={[pr]below:$0.294$}] (l57) at (0,-4.2) {57};
\node[sib,  label={[pr]below:$0.126$}] (l58) at (1.3,-4.2) {58};
\node[cand, label={[pr]below:$0.168$}] (l54) at (2.6,-4.2) {54};
\node[sib,  label={[pr]below:$0.112$}] (l55) at (3.9,-4.2) {55};
\node[cand, label={[pr]below:$0.180$}] (l14) at (5.2,-4.2) {14};
\node[sib,  label={[pr]below:$0.060$}] (l15) at (6.5,-4.2) {15};
\draw (root) -- node[ed] {0.7} (n12); \draw (root) -- node[ed] {0.3} (n13);
\draw (n12) -- node[ed] {0.6} (n201); \draw (n12) -- node[ed] {0.4} (n200);
\draw (n13) -- node[ed] {0.8} (n123); \draw[dashed] (n13) -- node[ed] {0.2} (n120);
\draw (n201) -- node[ed] {0.7} (l57); \draw (n201) -- node[ed] {0.3} (l58);
\draw (n200) -- node[ed] {0.6} (l54); \draw (n200) -- node[ed] {0.4} (l55);
\draw (n123) -- node[ed] {0.75} (l14); \draw (n123) -- node[ed] {0.25} (l15);
\node[pr, align=center] at (7.5,-3.45) {$p = 0.06$\\[-1pt](bound only)};
\node[lvl] at (-1.3,-1.4) {level 1}; \node[lvl] at (-1.3,-2.8) {level 2}; \node[lvl] at (-1.3,-4.2) {level 3};
\node[pr, align=center] at (4.3,-5.05) {top-2 $= \{57, 14\}$, $s_2 = 0.180 \ge U_0 = 0.06$: certified after one pass};
\begin{scope}[xshift=10.6cm]
\node[ttl] at (4.3,0.9) {(b) beam search, width $W = 2$, on the same request};
\node[root] (broot) at (4.3,0) {root ($h$)};
\node[kept] (b12) at (1.95,-1.4) {12};
\node[kept] (b13) at (6.675,-1.4) {13};
\node[kept, label={[pr]right:$0.42$}] (b201) at (0.65,-2.8) {201};
\node[kept, label={[pr]right:$0.28$}] (b200) at (3.25,-2.8) {200};
\node[pruned, label={[prp]right:$0.24$}] (b123) at (5.85,-2.8) {123};
\node[pruned, label={[prp]right:$0.06$}] (b120) at (7.5,-2.8) {120};
\node[ret, label={[pr]below:$0.294$}] (m57) at (0,-4.2) {57};
\node[scored, label={[pr]below:$0.126$}] (m58) at (1.3,-4.2) {58};
\node[ret, label={[pr]below:$0.168$}] (m54) at (2.6,-4.2) {54};
\node[scored, label={[pr]below:$0.112$}] (m55) at (3.9,-4.2) {55};
\node[miss, label={[prm]below:$0.180$}] (m14) at (5.2,-4.2) {14};
\node[pruned, label={[prp]below:$0.060$}] (m15) at (6.5,-4.2) {15};
\draw (broot) -- node[ed] {0.7} (b12); \draw (broot) -- node[ed] {0.3} (b13);
\draw (b12) -- node[ed] {0.6} (b201); \draw (b12) -- node[ed] {0.4} (b200);
\draw (b13) -- node[ed] {0.8} (b123); \draw (b13) -- node[ed] {0.2} (b120);
\draw (b201) -- node[ed] {0.7} (m57); \draw (b201) -- node[ed] {0.3} (m58);
\draw (b200) -- node[ed] {0.6} (m54); \draw (b200) -- node[ed] {0.4} (m55);
\draw[dashed, black!45] (b123) -- (m14); \draw[dashed, black!45] (b123) -- (m15);
\node[prp, align=center, anchor=west] at (8.75,-2.8) {beam keeps two:\\[-1pt]$0.24 < 0.28$, pruned};
\node[prm, align=center, anchor=west] at (7.25,-4.2) {$14$: the second-best\\[-1pt]item, never scored};
\node[lvl] at (-1.3,-1.4) {pass 1}; \node[lvl] at (-1.3,-2.8) {pass 2}; \node[lvl] at (-1.3,-4.2) {pass 3};
\node[pr, align=center] at (4.3,-5.05)
{returns $\{57, 54\}$ after three passes, and nothing tells it that $14$ is missing};
\end{scope}
\begin{scope}[yshift=-5.9cm, xshift=1.2cm, every node/.style={inner sep=1.5pt}, font=\scriptsize]
\node[fed, minimum width=5mm, minimum height=3.5mm] (lg1) at (0,0) {}; \node[anchor=west] at (lg1.east)
{\,expanded or kept};
\node[cand, minimum width=5mm, minimum height=3.5mm] (lg2) at (3.0,0) {}; \node[anchor=west] at (lg2.east)
{\,candidate / returned};
\node[sib, minimum width=5mm, minimum height=3.5mm] (lg3) at (6.3,0) {}; \node[anchor=west] at (lg3.east)
{\,free sibling, scored};
\node[front, minimum width=5mm, minimum height=3.5mm] (lg4) at (9.6,0) {}; \node[anchor=west] at (lg4.east)
{\,frontier, bound only};
\node[pruned, minimum width=5mm, minimum height=3.5mm] (lg5) at (12.6,0) {}; \node[anchor=west] at (lg5.east)
{\,pruned by the beam};
\node[miss, minimum width=5mm, minimum height=3.5mm] (lg6) at (15.6,0) {}; \node[anchor=west] at (lg6.east)
{\,item of \ModelK that beam search misses};
\end{scope}
\end{tikzpicture}}
\caption{\textbf{The same request under the two decoders} ($L = 3$, $K = 2$).
\textbf{(a)} One pass over the draft scores the candidates and their free siblings, and the only unexpanded prefix
	weighs less than the second-best score, so the first pass is certified.
\textbf{(b)} Beam search of width $K$ keeps the two heaviest prefixes per level and prunes $(13, 123)$, which holds the
	second-best item.
It returns a wrong list after three passes, and nothing in its computation reveals the error.}
\label{fig:example}
\end{figure*}
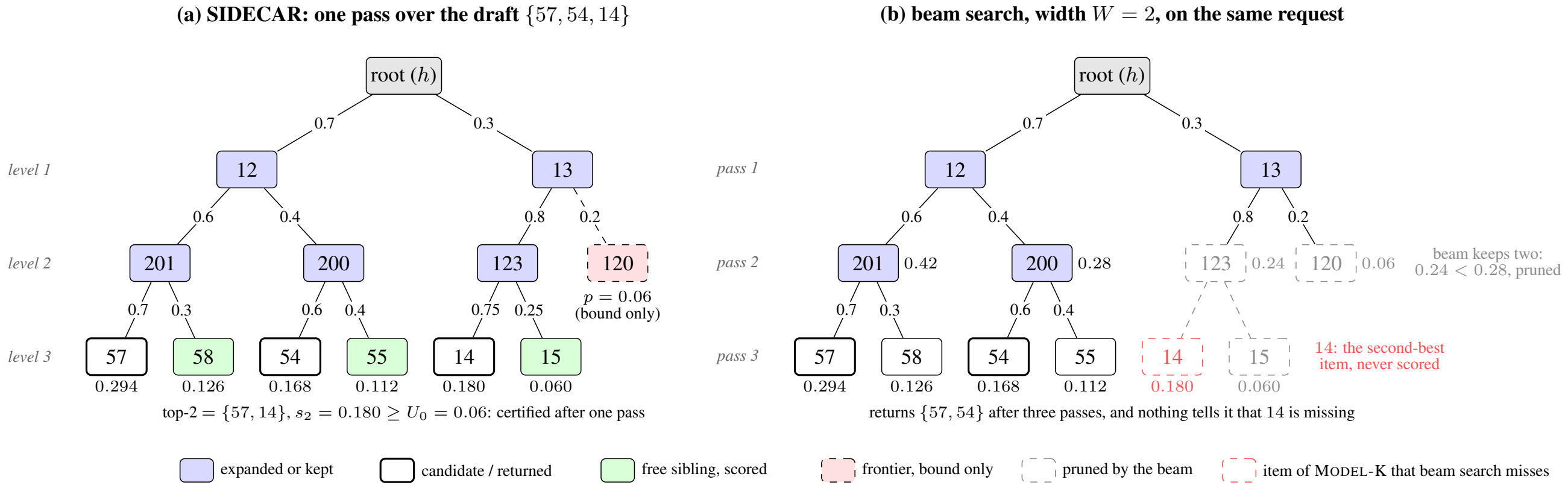

%% file: sections/02_decoder_and_certificate.tex
\section{The decoder and its certificate}
\label{sec:onepass}

In this section, we describe the setting of constrained decoding and provide the necessary notation.
Then, we present \method{} with unbounded memory per pass, namely its first pass, its certificate, and its fallback, and
	we show how the quality of the draft affects the cost.

\paragraph{Constrained decoding over the catalog trie.}
Throughout the paper, every item $e$ of the catalog $\cE$ has a Semantic ID $\sid(e) = (c_1(e), \dots, c_L(e))$, where
	the first $L-1$ codes come from the content of the item and the last code separates items with equal content codes.
The SIDs of the catalog form the \emph{catalog trie} $\cT$, which the serving stack holds.
Given the user history $h$ and a prefix $c_{<\ell}$, the logits $z$ of the model define, as in constrained decoding,
	probabilities over the admissible next codes only:
\beq{
p(c_\ell \mid h, c_{<\ell}) = \frac{\exp z_{c_\ell}}{\sum_{c \in \children(c_{<\ell})} \exp z_c},
\label{eq:constrained}
}
where $\children(c_{<\ell})$ denotes the admissible children of the prefix in $\cT$.
The probability of an item is the product of its $L$ conditionals, and the probability of a prefix $b$ is the product of
	the first $\depth(b)$ of them.
Since the conditionals at a prefix sum to one over its children, the probability of a prefix equals the total
	probability of the items below it.

Our target is the exact top-$K$ of the model, i.e., the $K$ items of highest probability under the model for the request
	at hand, which we denote by \ModelK.
Every decoder returns a list of $K$ items; constrained beam search returns \BeamK, the top-$K$ of its final beam after
	$L$ sequential passes with $W$ prefixes per level, which only approximates \ModelK.
Our decoder returns \SidecarK.
We use \SidecarKr{r} to denote its output list after round $r$ of the procedure below.
We call a returned list \emph{exact} when it equals \ModelK.
\SidecarK is \emph{within a factor} $1 + \delta$ when every item of \ModelK is within that factor of the item at its
	rank.

\paragraph{The first pass.}
A \emph{draft} is any set $C$ of $N$ candidate items.
We map the candidates to their SIDs and insert them into a trie, henceforth referred to as the \emph{candidate trie}.
The internal nodes follow the history in the input, and a tree-attention mask lets every node attend to the history, to
	its ancestors, and to itself.
With position ids that follow the depth of a node, the output at every node equals the output that a separate pass over
	its own path would produce.
Hence, one pass returns the model's distribution at every internal node of the candidate trie.
From these outputs, we read two kinds of values.
First, every item whose depth-$(L-1)$ prefix is in the trie receives its exact probability, which includes the
	candidates and also their siblings at no extra cost.
Second, every admissible child of an expanded node that is not itself expanded is a \emph{frontier} prefix, and the pass
	returns its probability as well, which bounds every item below it.
\Cref{fig:example} illustrates the pass on a small example.

After any round $r$, let $E_r$ denote the set of items with exact probabilities, $s_K(r)$ the $K$-th largest of these
	probabilities, $F_r$ the frontier, and $U_r = \max_{b \in F_r} p(b \mid h)$ the largest frontier probability.
The list \SidecarKr{r} is the top-$K$ of $E_r$.
Since every item outside $E_r$ lies below exactly one frontier prefix, its probability is at most $U_r$, which gives the
	following certificate.

\begin{theorem}[certificate]
\label{thm:cert}
If $s_K(r) \ge U_r$, then $\SidecarKr{r} = \ModelK$.
\end{theorem}

When the condition of \cref{thm:cert} holds, we call \SidecarK \emph{certified}.
Figures and tables name the two outcomes, exact and certified, \Exact and \Certificate respectively.
Note that \BeamK can be exact on some requests but never certified, because beam search cannot discern its exact lists
	from its other lists.

The test uses only quantities that the completed passes have already computed, namely a sort of the exact scores and a
	maximum over the frontier.
In other words, the draft decides only how soon the certification test passes.
If it contains the exact top-$K$, the test passes after the first pass.
Otherwise, the fallback rounds continue until the test passes, and the returned list is still the exact top-$K$.

\paragraph{The fallback.}
If the certification fails after the first pass, some frontier prefix may hide a better item underneath.
We call a frontier prefix \emph{open} when its probability exceeds the current $K$-th best score, and we denote the open
	set by $O_r = \{ b \in F_r : p(b \mid h) > s_K(r) \}$.
The next round only expands the open prefixes, in one pass that reuses the cached history and ancestors, and reads their
	children.
\Cref{alg:main} summarizes the procedure.

\begin{algorithm}[t]
\caption{\method{}}
\label{alg:main}
\begin{algorithmic}[1]
\REQUIRE history $h$, draft $C$, list size $K$
\STATE $E_0, F_0 \gets$ one tree-attention pass over the candidate trie of $C$ after $h$
\FOR{$r = 0, 1, \dots, L-2$}
  \STATE $O_r \gets \{ b \in F_r : p(b \mid h) > s_K(r) \}$
  \IF{$O_r = \emptyset$} \STATE \textbf{return} top-$K$ of $E_r$ \ENDIF
  \STATE $E_{r+1}, F_{r+1} \gets$ one pass over the children of $O_r$ with the cached KV
\ENDFOR
\STATE \textbf{return} top-$K$ of $E_{L-1}$
\end{algorithmic}
\end{algorithm}

\begin{proposition}[termination]
\label{prop:anydraft}
Regardless of the draft and of the model, $s_K$ never decreases and $U$ never increases from one round to the next, and
	\cref{alg:main} returns \ModelK after at most $L$ passes.
\end{proposition}

Intuitively, each round descends one level of the trie below the open prefixes, so after $L-1$ rounds nothing can remain
	open.
Thus, a good draft lets the procedure stop after the first pass.
A poor draft requires more rounds, but the list is still exact.

\paragraph{Early stopping before the certificate holds.}
A deployment may stop the procedure before the certificate holds, for instance at a cap on the number of passes that fit
	under the maximum serving latency.
The list at that point then carries a weaker but valid statement.
Let $s_k(r)$ denote the $k$-th highest exact score and $p^*_{(k)}$ the probability of the model's $k$-th best item, and
	define the \emph{gap} as $\gap_r = U_r / s_K(r) - 1$.

\begin{proposition}[per-rank bound]
\label{prop:gap}
After any round $r$ and for every rank $k \le K$,
\begin{align*}
s_k(r) \le p^*_{(k)} &\le \max\bigl(s_k(r), U_r\bigr) \\
&\le \bigl(1 + \max(\gap_r, 0)\bigr)\, s_k(r).
\end{align*}
\end{proposition}

Hence, \SidecarKr{r} is within a factor $1 + \gap_r$ of \ModelK at every rank, and $\gap_r \le 0$ is exactly the
	certificate.
Note that beam search offers neither statement, since it keeps no bound on the items it prunes.

\paragraph{Effect of draft quality.}
The quality of the draft never affects exactness.
Instead, it affects the number of passes, and it does so through one quantity, namely the probability that the first
	pass leaves unscored, $\varepsilon = 1 - p(E_0)$, which we call the \emph{hidden probability}.
If $\varepsilon < p^*_{(K)}$, i.e., if the hidden probability is below the probability of the model's $K$-th best item,
	then \SidecarKr{0} is certified.
To see why, note that the frontier prefixes of the first pass are disjoint and hold exactly the unscored probability, so
	each of them has probability at most $\varepsilon$.
An unscored item of \ModelK would lie below one of them with probability at least $p^*_{(K)} > \varepsilon$, which is
	impossible.
Hence, every item of \ModelK is scored, $s_K(0) = p^*_{(K)} > \varepsilon \ge U_0$, and the certificate holds.
Thus, a draft does not have to contain \ModelK.
It only has to contain items from the right prefixes, so that the probability it leaves uncovered is smaller than that
	of the $K$-th best item.

%% file: sections/03_catalog_trie_structure.tex
\section{Exploiting the structure of the catalog trie}
\label{sec:trie}

Building upon the decoder of \cref{sec:onepass}, this section shows what the structure of the catalog trie contributes
	to \method{}.
First, we observe that most deep prefixes of a SID trie hold a single item.
Then, we use this structure to save tokens in every pass and to resolve some open prefixes in a single round.

\paragraph{Single-item prefixes.}
In a SID trie, most deep prefixes hold a single item, and this follows from the size of the code space rather than from
	any one catalog or quantizer.
The $L-1$ content codes, drawn from codebooks of size $B$, address $B^{L-1}$ cells, which is orders of magnitude more
	than the thousands to millions of items that a catalog holds (e.g., 16.7 million cells with the three codebooks of
	256 codes of TIGER~\citep{rajput2023tiger}).
Moreover, every quantizer in use, whether a residual quantizer~\citep{lee2022rqvae,deng2025onerec} or a hierarchical
	clustering of the item embeddings~\citep{tay2022dsi,wang2022nci}, spreads the items of a prefix over its codebook at
	every level, so that the number of items per prefix falls geometrically with depth.
Indeed, the collision code of TIGER exists because the content codes alone leave almost every item in a prefix of its
	own~\citep{rajput2023tiger}.
Thus, below the first level, most prefixes hold a single item.
At such a prefix, the model has one admissible child, so its constrained conditional is one and the item below inherits
	the probability of the prefix.
Consequently, these prefixes need no pass and no token.

We call a prefix \emph{branching} when it has at least two admissible children.
The \emph{skeleton} of a prefix $b$, denoted by $S(b)$, is the set of branching prefixes below $b$ together with the
	non-branching prefixes on the paths that connect them to $b$, which we call \emph{connectors}.
Note that the connectors are needed because a node can only be fed together with its ancestors.
We use $\sigma(b) = |S(b)|$ to denote the number of nodes in the skeleton, i.e., the number of tokens that a pass has to
	feed to resolve every item below $b$ at once.
Both $S(b)$ and $\sigma(b)$ depend only on the catalog trie, so they are known before any request arrives, and the
	outputs at the skeleton determine the probability of every item below $b$.
Since most deep prefixes are chains, the skeleton of a cluster is a small fraction of its prefixes, and the skeleton of
	the whole catalog is a small fraction of the trie.

\paragraph{Chain rules.}
We exploit the chains in three places of \method{}, and in each of them we feed only nodes that are known to be needed.
First, in the first pass, a candidate prefix whose subtree is a chain contributes nothing to the scores, because the
	conditional at its parent gives its probability, and every node on the chain passes this probability unchanged to
	the item.
Hence, the first pass feeds only the branching candidate prefixes and the connectors on the candidate paths.
Second, a frontier prefix whose subtree is a chain is resolved without a token for the same reason, since its item
	receives the probability of the prefix, which the pass has already returned.
Third, an open prefix with a single admissible child makes that child open too, since the child inherits its
	probability.
Thus, a fallback round feeds the whole unary run down to the next branching prefix at once, instead of one node per
	round.
These rules leave the certificate, the exactness, and the bound of $L$ passes untouched.

\paragraph{The skeleton round.}
A fallback round of \cref{alg:main} feeds the open prefixes and reads the conditionals of their children, which become
	the frontier of the next round.
Hence, the fallback descends one level per round, and this is where the bound of $L-1$ rounds comes from.
When the memory per pass is unbounded, the tree pass can instead feed an open prefix together with its whole skeleton,
	exactly as the first pass feeds the whole candidate trie.
The ancestors of every fed node are then in the sequence, as tree attention requires, and consequently every item below
	the open prefix has an exact probability after the pass.
Hence, if the first fallback round feeds every open prefix $b \in O_0$ together with its skeleton, at a cost of $\sum_{b
	\in O_0} \sigma(b)$ tokens, nothing below $O_0$ remains on the frontier, and every other frontier prefix stays
	closed, since $s_K$ does not decrease.
Thus, the open set is empty after the second pass, and \SidecarK is certified, regardless of the draft and of the model.

In other words, the skeleton round saves rounds at the cost of additional tokens.
Under a bound on the nodes of a pass, however, the skeletons of all open prefixes may not fit into one pass.
In the memory-bound regime of serving, a pass of a few thousand tokens costs little more than a pass of fifty, so the
	trade favors the skeleton whenever the open set is a moderate number of clusters.
Even in the worst case over all drafts, with every cluster open, the round feeds the skeleton of the catalog, which is a
	fraction of the nodes that an exhaustive pass over the trie would feed.
Unlike the chain rules, the skeleton round feeds nodes before knowing that they are open, so it is not part of the
	default decoder.

\input{sections/fig_plans}   

%% file: sections/fig_plans.tex
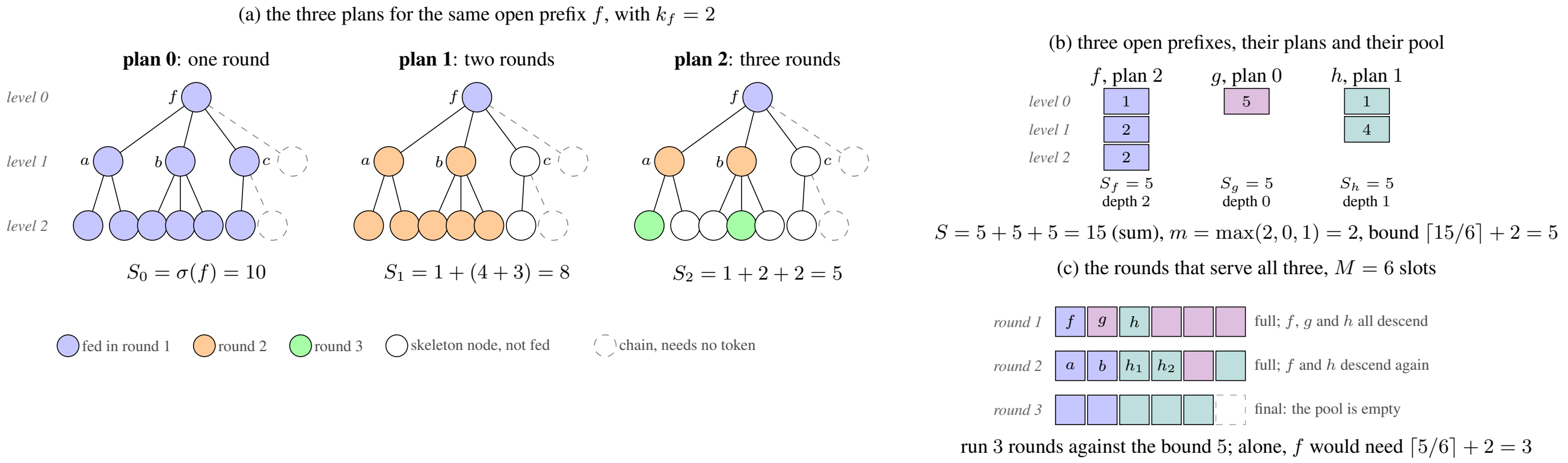
\begin{figure*}[t]
\centering
\resizebox{\textwidth}{!}{%
\begin{tikzpicture}[x=1.25cm, y=1.25cm,
  font=\footnotesize, every node/.style={inner sep=1.5pt},
  nd/.style={draw, circle, minimum size=4.6mm, inner sep=0pt, font=\scriptsize},
  r1/.style={nd, fill=blue!22}, r2/.style={nd, fill=orange!40}, r3/.style={nd, fill=green!35},
  no/.style={nd, fill=white}, ch/.style={nd, dashed, fill=white, draw=black!50},
  lvl/.style={font=\scriptsize\itshape, text=black!60, anchor=east},
  slot/.style={draw, minimum width=4.6mm, minimum height=4.6mm, inner sep=0pt, font=\scriptsize},
  sf/.style={slot, fill=blue!22}, sg/.style={slot, fill=violet!25}, se/.style={slot, fill=white, draw=black!35, dashed},
  sh/.style={slot, fill=teal!25},
  pb/.style={draw, minimum width=7mm, minimum height=4mm, inner sep=0pt, font=\scriptsize},
  ttl/.style={font=\footnotesize, align=center},
  ann/.style={font=\scriptsize, text=black!70, anchor=west}]
\newcommand{\treeskel}[2]{%
\begin{scope}[xshift=#1*1.25cm]
\coordinate (f) at (0,0);
\coordinate (a) at (-1.1,-0.8); \coordinate (b) at (-0.2,-0.8); \coordinate (c) at (0.6,-0.8); \coordinate (d) at
(1.2,-0.8);
\coordinate (a1) at (-1.35,-1.6); \coordinate (a2) at (-0.9,-1.6);
\coordinate (b1) at (-0.55,-1.6); \coordinate (b2) at (-0.2,-1.6); \coordinate (b3) at (0.15,-1.6);
\coordinate (c1) at (0.55,-1.6); \coordinate (c2) at (0.95,-1.6);
\draw (f)--(a) (f)--(b) (f)--(c) (a)--(a1) (a)--(a2) (b)--(b1) (b)--(b2) (b)--(b3) (c)--(c1);
\draw[dashed, black!50] (f)--(d) (c)--(c2);
\foreach \n/\s in {#2} {\node[\s] at (\n) {};}
\node[ch] at (d) {}; \node[ch] at (c2) {};
\node[font=\scriptsize] at ([xshift=-3.6mm]f) {$f$};
\node[font=\scriptsize] at ([xshift=-3.6mm]a) {$a$}; \node[font=\scriptsize] at ([xshift=-3.4mm]b) {$b$};
\node[font=\scriptsize] at ([xshift=3.4mm]c) {$c$};
\end{scope}}
\node[lvl] at (-1.8,0) {level 0}; \node[lvl] at (-1.8,-0.8) {level 1}; \node[lvl] at (-1.8,-1.6) {level 2};
\treeskel{0}{f/r1, a/r1, b/r1, c/r1, a1/r1, a2/r1, b1/r1, b2/r1, b3/r1, c1/r1}
\treeskel{3.5}{f/r1, a/r2, b/r2, c/no, a1/r2, a2/r2, b1/r2, b2/r2, b3/r2, c1/no}
\treeskel{7}{f/r1, a/r2, b/r2, c/no, a1/r3, a2/no, b1/no, b2/r3, b3/no, c1/no}
\node[ttl] at (0,-2.2) {$S_0 = \sigma(f) = 10$};
\node[ttl] at (3.5,-2.2) {$S_1 = 1 + (4 + 3) = 8$};
\node[ttl] at (7,-2.2) {$S_2 = 1 + 2 + 2 = 5$};
\node[ttl, anchor=south] at (0,0.3) {\textbf{plan 0}: one round}; \node[ttl, anchor=south] at (3.5,0.3)
{\textbf{plan 1}: two rounds};
\node[ttl, anchor=south] at (7,0.3) {\textbf{plan 2}: three rounds};
\node[ttl, anchor=south] at (3.5,0.85) {(a) the three plans for the same open prefix $f$, with $k_f = 2$};
\begin{scope}[yshift=-3.1*1.25cm, every node/.style={inner sep=1pt}]
\node[r1, minimum size=3.4mm] (l1) at (-1.6,0) {}; \node[ann] at (l1.east) {fed in round 1};
\node[r2, minimum size=3.4mm] (l2) at (0.1,0) {}; \node[ann] at (l2.east) {round 2};
\node[r3, minimum size=3.4mm] (l3) at (1.3,0) {}; \node[ann] at (l3.east) {round 3};
\node[no, minimum size=3.4mm] (l4) at (2.5,0) {}; \node[ann] at (l4.east) {skeleton node, not fed};
\node[ch, minimum size=3.4mm] (l5) at (5.1,0) {}; \node[ann] at (l5.east) {chain, needs no token};
\end{scope}
\newcommand{\slotrow}[4]{%
\begin{scope}[xshift=#1*1.25cm, yshift=#2*1.25cm]
\foreach \s/\l [count=\i from 0] in {#3} {\node[\s] at (\i*0.4,0) {\l};}
\node[ann] at (2.25,0) {#4};
\end{scope}}
\begin{scope}[xshift=10.9*1.25cm]
\node[ttl, anchor=south] at (2.2,0.5) {(b) three open prefixes, their plans and their pool};
\node[lvl] at (0.05,-0.05) {level 0}; \node[lvl] at (0.05,-0.4) {level 1}; \node[lvl] at (0.05,-0.75) {level 2};
\node[ttl] at (0.7,0.25) {$f$, plan $2$}; \node[ttl] at (2.2,0.25) {$g$, plan $0$}; \node[ttl] at (3.7,0.25)
{$h$, plan $1$};
\node[pb, fill=blue!22] at (0.7,-0.05) {$1$}; \node[pb, fill=blue!22] at (0.7,-0.4) {$2$}; \node[pb, fill=blue!22] at
(0.7,-0.75) {$2$};
\node[pb, fill=violet!25] at (2.2,-0.05) {$5$};
\node[pb, fill=teal!25] at (3.7,-0.05) {$1$}; \node[pb, fill=teal!25] at (3.7,-0.4) {$4$};
\node[ttl, font=\scriptsize] at (0.7,-1.2) {$S_f = 5$\\depth $2$}; \node[ttl, font=\scriptsize] at (2.2,-1.2)
{$S_g = 5$\\depth $0$}; \node[ttl, font=\scriptsize] at (3.7,-1.2) {$S_h = 5$\\depth $1$};
\node[ttl] at (2.2,-1.7) {$S = 5 + 5 + 5 = 15$ (sum), $m = \max(2, 0, 1) = 2$, bound $\lceil 15/6 \rceil + 2 = 5$};
\node[ttl, anchor=south] at (2.2,-2.3) {(c) the rounds that serve all three, $M = 6$ slots};
\node[lvl] at (-0.3,-2.8) {round 1};
\slotrow{0}{-2.8}{sf/$f$, sg/$g$, sh/$h$, sg/, sg/, sg/}{full; $f$, $g$ and $h$ all descend}
\node[lvl] at (-0.3,-3.35) {round 2};
\slotrow{0}{-3.35}{sf/$a$, sf/$b$, sh/$h_1$, sh/$h_2$, sg/, sh/}{full; $f$ and $h$ descend again}
\node[lvl] at (-0.3,-3.9) {round 3}; \slotrow{0}{-3.9}{sf/, sf/, sh/, sh/, sh/, se/}{final: the pool is empty}
\node[ttl, anchor=north] at (2.2,-4.2)
{run $3$ rounds against the bound $5$; alone, $f$ would need $\lceil 5/6 \rceil + 2 = 3$};
\end{scope}
\end{tikzpicture}}
\caption{\textbf{Plans and rounds under a budget} (\cref{prop:plancost,prop:rounds}).
\textbf{(a)} The three plans of one open prefix with $k_f = 2$, so that at most two nodes per level can be open: plan
	$0$ feeds the whole skeleton in one round, plan $2$ one level per round, plan $1$ lies between, and the adversary
	opens the nodes with the largest skeletons.
\textbf{(b)} Three open prefixes with different plans: the pool size $S$ adds over the prefixes, whereas the depth $m$
	is their maximum, since they descend together.
\textbf{(c)} Six slots per round serve all three in three rounds, known-open prefixes first and skeleton nodes after,
	which beats the bound of five because no round was under-filled while depth remained.}
\label{fig:plans}
\end{figure*}

%% file: sections/04_decoding_under_budget.tex
\section{Decoding under a budget}
\label{sec:budgets}

\begin{algorithm}[t]
\caption{\bmethod{}: \method{} with $M$ nodes per pass, $Y$ passes, and a target factor $1 + \delta_t$ set by the
	deployment}
\label{alg:budget}
\begin{algorithmic}[1]
\REQUIRE history $h$, draft $C$, size $K$, target factor $1 + \delta_t$ (default $\delta_t = 0$, i.e., exact)
\STATE \algline{ln:first} $E_0, F_0 \gets$ first pass over the candidate trie of $C$: at most $M$ skeleton members, in
draft order
\STATE \algline{ln:announce} \textbf{announce} the factor $1 + \max(\delta^*(M, Y), \delta_t)$ for $Y$ passes, and
$Y_{\min}$ passes for exactness
\FOR{$r = 0, 1, \dots, Y-2$}
  \STATE \algline{ln:stop} \textbf{if} no prefix of $F_r$ is open \textbf{then return} top-$K$ of $E_r$
  \STATE \algline{ln:replan} re-plan from $F_r$: factor in force $\Delta_r$, pool, plans and schedule $x_r$
  \STATE \algline{ln:batch} $B \gets$ the heaviest open prefixes, at most $M$
  \STATE \algline{ln:spec} add pool nodes until $B$ holds $x_r$ of them, displacing the lightest prefixes outside the
  pool, then fill the free slots with more
  \STATE \algline{ln:pass} $E_{r+1}, F_{r+1} \gets$ one pass over $B$ with the cached KV
\ENDFOR
\STATE \textbf{return} top-$K$ of $E_{Y-1}$
\end{algorithmic}
\end{algorithm}

So far, we have assumed that a pass can hold as many nodes as the draft and the fallback need.
In practice, however, a serving deployment provisions its resources in advance and applies the same limits to every
	request.
It bounds the memory of each pass by the number of nodes that a pass may hold, denoted by $M$, and the latency of each
	request by the maximum number of passes that it may run, denoted by $Y$.
Note that for $Y = L$ these are exactly the resources of constrained beam search with width $W = M$.
In this section, we address two questions about such a budget.
First, what can \method{} guarantee when the budget may be too small for exactness?
Second, how soon can we provide this guarantee, i.e., before spending the remaining budget?
The guarantee takes the form of an approximation factor: \SidecarK is within a factor $1 + \delta$ of \ModelK at every
	rank, in the sense of \cref{prop:gap}.
We call the smallest $\delta$ that can be stated after the first pass the \emph{announced factor}.
We refer to \method{} under such a budget as \bmethod{} (\cref{alg:budget}), and to its execution as the \emph{planned
	run}.
First, we describe what the first pass leaves us with (line~\ref{ln:first} of \cref{alg:budget}).
Second, we bound what resolving one open prefix can cost.
Third, we show how the costs of all open prefixes translate into a number of rounds and into a plan for each round
	(lines~\ref{ln:replan}--\ref{ln:spec}).
Finally, we derive the factor that the rounds within the budget can guarantee in the worst case
	(line~\ref{ln:announce}).

Throughout this section, we follow our established conventions.
First, the state after round $r$ carries the index $r$, i.e., $E_r$ is the set of scored items, $s_K(r)$ the $K$-th best
	exact score, and $F_r$ the frontier.
Round $0$ is the first pass.
Hence, after round $r$, the planned run has used $1 + r$ passes, and $Y - 1 - r$ fallback rounds remain.
We state the construction below for $r = 0$, which is the moment of the announcement, and line~\ref{ln:replan} of
	\cref{alg:budget} repeats it after every round.
Second, the guarantee is a worst case over what the first pass has not observed.
After the first pass, the decoder knows the outputs of the model at the fed nodes, but not how the probability of each
	frontier prefix splits among the prefixes below it.
We call any assignment of these unobserved outputs that agrees with the first pass a \emph{consistent completion}, and
	we let an \emph{adversary} choose the completion that hurts the fallback most.
The announced factor holds against this adversary, and consequently for every consistent completion.
Since the real model is one of the consistent completions, the factor holds for it as well, usually with much room to
	spare.

\paragraph{Threshold and open prefixes.}
The first pass (line~\ref{ln:first} of \cref{alg:budget}) is the first pass of \method{} capped at $M$ nodes, i.e., it
	feeds the skeleton members of the candidate trie in draft order for as long as they fit.
It returns the scored items $E_0$, their $K$-th best score $s_K(0)$, and the frontier $F_0$, together with the
	probability $p(f)$ of every frontier prefix $f$.
Below the frontier, however, nothing is yet known.
From these quantities, we can determine whether the current list is within a given factor of \ModelK.
Every unscored item lies below some frontier prefix, so its probability is at most that of the prefix.
Hence, \SidecarK is within the factor $1 + \delta$ as soon as every frontier prefix $b$ satisfies $p(b) \le (1 +
	\delta)\, s_K$, where $s_K$ is the current $K$-th best exact score.
Let $t = (1 + \delta)\, s_K(0)$ denote this \emph{threshold} after the first pass.
We call a frontier prefix \emph{open} at $t$ when $p(b) > t$.
Since $s_K$ never decreases, a prefix that is closed at $t$ stays closed.
Thus, the task of the fallback is to bring every open prefix below $t$ within the allowed rounds and nodes.
Note that the announcement is made before the fallback runs, so it can use only the quantities returned by the first
	pass and the catalog trie.

The skeleton of a frontier prefix $b$ also tells how many levels of expansion resolve it under the one-level-per-round
	fallback, namely one plus the depth of its deepest branching prefix minus the depth of $b$.
We call this number the \emph{resolution depth} $d_b \le L - \depth(b)$.

\paragraph{Plans for one open prefix.}
Fix an open prefix $f \in F_0$.
The fallback has to find out whether some prefix below $f$ is also above $t$, and it has to keep doing so until no such
	prefix remains.
We count \emph{levels} from $f$: $f$ itself is at level $0$, its children are at level $1$, and its descendants $j$
	levels down are at level $j$.
Two facts bound the number of prefixes at level $j$ that can be above $t$, and both of them are known after the first
	pass.
First, the prefixes at any one level below $f$ are disjoint and have total probability $p(f)$, so at most $k_f =
	\floor{p(f)/t}$ of them can exceed $t$.
Second, only skeleton nodes need to be fed (\cref{sec:trie}), and only $c_j(f)$ of them exist at level $j$, where
	$c_0(f) = 1$ counts $f$ itself and $\sigma(f) = \sum_j c_j(f)$ is the size of the skeleton of $f$.
Within these two bounds, however, nothing observed so far tells which nodes are open, so we let the adversary decide.

The fallback can resolve $f$ in two ways.
The first way is \emph{level by level}: the fallback feeds $f$, reads its children, feeds the open ones among them, and
	so on, descending one level per round.
Every node fed in this way is known to be open.
However, each level requires a separate round, so resolving $f$ takes up to $d_f$ rounds, where $d_f$ is its resolution
	depth.
The second way is the \emph{whole skeleton}: the fallback feeds $f$ together with its entire skeleton in one round.
This takes a single round, but it costs $\sigma(f)$ nodes whether or not they turn out to be open.
Note that feeding only part of a level does not help against the adversary, who can place the hidden probability on
	whatever part of the level was skipped.
Therefore, the options that admit a worst-case guarantee are the mixtures of the two ways, in which the fallback
	descends $\ell$ levels one round at a time and then feeds the open nodes of level $\ell$ together with their whole
	skeletons.
We call $\ell_f \in \{0, \dots, L - 1 - \depth(f)\}$ the \emph{plan} of $f$, and \cref{fig:plans}(a) shows the three
	plans of one open prefix.

\begin{proposition}[worst-case cost of a plan]
\label{prop:plancost}
Let $f$ be open at threshold $t$ and $k_f = \floor{p(f)/t}$.
For every consistent completion:
\begin{enumerate}[label=\textup{(\roman*)}]
\item at most $\min(c_j(f), k_f)$ nodes at level $j$ below $f$ are above $t$ and have to be fed;
\item the plan $\ell$ feeds at most
\beq{
S_\ell(f, t) \;=\; \underbrace{\sum_{j=0}^{\ell-1} \min\bigl(c_j(f), k_f\bigr)}_{\text{levels fed one at a time}}
\;+\; \underbrace{\Sigma^{\sigma}_{\ell}(f, t)}_{\text{skeletons at level } \ell}
\label{eq:plan}
}
nodes before nothing below $f$ is open, where $\Sigma^\sigma_{\ell}(f, t)$ is the sum of the $k_f$ largest skeleton
	sizes among the nodes at level $\ell$.
\end{enumerate}
\end{proposition}

The plan $\ell = 0$ is the skeleton round, at a cost of $\sigma(f)$ nodes.
The largest plan is pure level-by-level descent.
It feeds only nodes that are known to be open, but it needs one round per level.
Between these two extremes, a larger plan uses more rounds and feeds fewer nodes.

\paragraph{Combining the plans of all open prefixes.}
Suppose that every open prefix has been given a plan, and let the \emph{pool} denote the set of nodes that these plans
	may feed.
The pool has a worst-case size $S$, namely the sum of \cref{eq:plan} over the open prefixes, and a depth $m = \max_f
	\ell_f$, i.e., the largest number of levels that any prefix descends one round at a time.
Note that both $S$ and $m$ are fixed by the plans and the catalog before any fallback round runs.
The planned run will discover which nodes are open.
However, since a closed prefix stays closed, every prefix that becomes open later is a descendant of some $f \in F_0$,
	and the plan of $f$ already provides for it.

\begin{proposition}[round bound and schedule]
\label{prop:rounds}
Let the open prefixes of $F_0$ have plans with pool size $S$ and depth $m$.
\begin{enumerate}[label=\textup{(\roman*)}]
\item The fallback finishes within
\beq{
\ceil{S / M} + m
\label{eq:rounds}
}
rounds, whatever the order in which full and under-filled rounds occur.
\item If $\ceil{S/M} + m \le y$ with $y$ rounds remaining, then
\beq{
x_r \;=\; S - (y - 1 - m)\, M \;\le\; M ,
\label{eq:schedule}
}
and a round that feeds at least $x_r$ pool nodes, or every pool node that is available, leaves a pool with $\ceil{S'/M}
	+ m' \le y - 1$.
\end{enumerate}
\end{proposition}

This bound counts two kinds of rounds.
A round that fills all its $M$ slots with pool nodes is \emph{full}, and since the plans feed at most $S$ pool nodes in
	total, there are at most $\ceil{S/M}$ full rounds.
A round that cannot be filled has fed every pool node whose parent is already known.
Hence, the depth that remains to be descended drops by one, and this can happen at most $m$ times.
Note that the two terms aggregate over the open prefixes in different ways.
Nodes are not shared between open prefixes, so their counts add up to $S$, whereas rounds are shared, since all open
	prefixes descend in the same rounds.
Hence, the depth term is the maximum $m$ of the plans rather than their sum.
\Cref{fig:plans}(b) and (c) show three open prefixes with different plans and the rounds that serve them together.

Part (ii) of \cref{prop:rounds} gives the rule for a single round, and lines~\ref{ln:batch} and~\ref{ln:spec} of
	\cref{alg:budget} carry it out.
We call the number $x_r$ the \emph{schedule}, i.e., the number of pool nodes that the current round must feed so that
	the round bound still fits into the rounds that remain after this one.
The pool nodes that are available in a round are of two kinds.
The first kind are the open prefixes themselves.
Every exact decoder has to feed them, so line~\ref{ln:batch} takes the heaviest of them that fit into the $M$ slots.
The second kind are the skeleton nodes below the prefixes with plan zero.
At this point, nothing tells whether they are needed, so feeding them prepares for the worst case without knowing
	whether it occurs.
Line~\ref{ln:spec} feeds such nodes only as far as the schedule requires: it adds them until the batch holds $x_r$ pool
	nodes, and when the batch is already full, they replace the lightest of its prefixes that lie outside the pool.
We call these nodes the \emph{speculative nodes}, and they are the only nodes that the planned run ever feeds without
	knowing that they are needed.
Finally, any slots that remain free are filled with further speculative nodes, since a pass takes about the same time
	whether or not its slots are filled, whereas a filled slot can resolve part of an open cluster.

\paragraph{The announced factor.}
The best plans are those that minimize the bound of \cref{eq:rounds}, and this minimization decomposes over the open
	prefixes.
To see why, fix a cap $m$ on the depth.
Then the depth term of the bound is at most $m$ for any plans with $\ell \le m$, so every open prefix can independently
	take its cheapest plan with $\ell \le m$.
For example, a heavy prefix with a small skeleton is fed whole, whereas a light prefix with a large skeleton descends
	level by level, and both happen in the same batch.
Let $S^*_m(t) = \sum_{f \in F_0 :\, p(f) > t} \min_{\ell \le m} S_\ell(f, t)$ denote the resulting pool size.
Minimizing over the cap then gives the smallest bound,
\beq{
R^*(t) \;=\; \min_{0 \le m \le L-2} \bigl( \ceil{S^*_m(t) / M} + m \bigr).
\label{eq:rstar}
}
Note that everything in $R^*(t)$ is known after the first pass: the frontier probabilities give $k_f$, the catalog gives
	$c_j(f)$ and the skeleton sizes, and $M$ is part of the budget.
Moreover, a larger threshold opens fewer prefixes and gives them smaller counts $k_f$, so $R^*$ is non-increasing in
	$t$.
Hence, the announced factor is the smallest $\delta$ for which these rounds fit into the $Y - 1$ fallback rounds of the
	budget,
\beq{
\delta^*(M, Y) \;=\; \min\bigl\{\delta \ge 0 : R^*\bigl((1+\delta)\, s_K(0)\bigr) \le Y - 1\bigr\},
\label{eq:deltaplan}
}
and Appendix~\ref{app:compute} explains how to compute it, either by bisection or exactly.
Similarly, we write $Y_{\min}(M) = 1 + R^*(s_K(0))$ for the smallest $Y$ with $\delta^*(M, Y) = 0$, i.e., the number of
	passes after which \SidecarK is certified with $M$ nodes per pass.
Both $\delta^*(M, Y)$ and $Y_{\min}(M)$ are known after the first pass, and line~\ref{ln:announce} of \cref{alg:budget}
	states them.

\paragraph{The planned run.}
A deployment may also hold a \emph{target} $\delta_t$, i.e., the factor that is good enough for it, with $\delta_t = 0$
	for exactness.
Two thresholds then govern each round.
The first is the target threshold $(1 + \delta_t)\, s_K(r)$: a frontier prefix is \emph{open} when it lies above this
	threshold, because it can then still change \SidecarK beyond what the target allows.
Hence, line~\ref{ln:stop} of \cref{alg:budget} stops only when no prefix is open, and line~\ref{ln:batch} feeds the open
	prefixes first.
The second is the threshold of the \emph{factor in force} $\Delta_r = \max(\delta^*_r, \delta_t)$, where $\delta^*_r$
	denotes \cref{eq:deltaplan} recomputed from $F_r$ and $s_K(r)$ with $Y - 1 - r$ rounds remaining, so that
	$\delta^*_0 = \delta^*(M, Y)$ is the announcement.

In each round, line~\ref{ln:replan} computes the pool, the plans, and the schedule only for the prefixes above $(1 +
	\Delta_r)\, s_K(r)$, since only these prefixes can violate the factor in force.
When $\Delta_r > \delta_t$, the prefixes between the two thresholds are open but outside the pool: line~\ref{ln:batch}
	feeds them when they are among the heaviest, and these are the prefixes that line~\ref{ln:spec} may displace.
When several sets of plans attain the same factor, the planned run uses the one with the smallest $S + mM$, which by
	\cref{eq:schedule} forces the least speculation.
Finally, line~\ref{ln:pass} is a fallback pass of \cref{alg:main}, with the batch in place of the open set.

\paragraph{Comparison with the unplanned run.}
In order to see what the planning changes, consider a run without it.
Without planning, \method{} under a budget is \cref{alg:main} with $M$ nodes per pass: in every round, it expands the
	heaviest open prefixes that fit, one level per round, and it stops when the certificate holds or when all $Y$ passes
	have been used.
We call this run the \emph{unplanned run}.
It never fails, in the sense that, by \cref{prop:gap}, the list that it holds at the end is within the factor $1 + \gap$
	at every rank.
However, when $Y$ passes are too few for the memory $M$, that list is not exact, its factor is whatever $\gap$ turns out
	to be, and nothing about this factor is known before the last pass.

In contrast, a factor that is stated after the first pass has to hold against the adversary, so the remaining passes
	must cover the worst case.
In particular, if the remaining rounds cannot resolve a heavy open prefix one level at a time, its skeleton has to be
	fed now, before the model has revealed whether it is needed.
Thus, the two runs differ only in line~\ref{ln:spec} of \cref{alg:budget}.
Both expand the prefixes that are known to be open, in the same order, as long as they fit.
In addition, \bmethod{} feeds the speculative nodes.
It first places them in the slots that the open prefixes leave empty, so that no open prefix is delayed.
Only when the schedule requires more speculative nodes than there are empty slots do they take the place of the lightest
	open prefixes outside the pool, which then wait for a later round.

Because the speculative nodes change what is known after each round, the two runs do not expand identical nodes in later
	rounds.
Nevertheless, the lists that they return within a fixed budget differ little, because the speculative nodes are what the
	adversary would need, and they rarely hold an item that the real model needs.
The planning adds two things.
First, the deployment knows the factor and $Y_{\min}$ after the first pass.
Second, when the passes are not capped, the certificate arrives sooner, since feeding a skeleton whole resolves a
	cluster in one round, where the level-by-level fallback needs several.
The difference is sharpest in the worst case, which the unplanned run does not cover.
When a heavy cluster hides the items of \ModelK across its skeleton, the level-by-level descent reaches those items only
	one level per round, so it can exhaust a budget that suffices for exactness, whereas \bmethod{} feeds the skeleton
	whole and is certified within the same budget, as the following proposition states.

\begin{proposition}[worst-case planning is necessary]
\label{prop:separation}
There exist a catalog trie, a draft, a model and a budget $(M, Y)$ with $Y < L$ under which the unplanned run, i.e.,
	\method{} (\cref{alg:main}) with $M$ nodes per pass, returns a list that contains no item of \ModelK, whereas
	\bmethod{} announces exactness after the first pass and returns \ModelK with its certificate within the same budget.
\end{proposition}

The instance in the proof is the worst case for the unplanned run, i.e., a model whose probability sits in one heavy
	cluster and is spread evenly over the items below it.
On this instance, only the planning lets the budget suffice for exactness.

\begin{theorem}[announced factor under a budget]
\label{thm:plan}
Run \bmethod{} with $M$ nodes per pass for at most $Y$ passes.
Then:
\begin{enumerate}[label=\textup{(\roman*)}]
\item \SidecarK is within a factor $1 + \delta^*(M, Y)$ of \ModelK at every rank;
\item $Y_{\min}(M) = 1 + R^*(s_K(0))$ passes certify \SidecarK, a number known after the first pass;
\item $\delta^*_{r+1} \le \delta^*_r$ for every round $r$: re-planning never loosens the announcement.
\end{enumerate}
\end{theorem}

\begin{remark}[\textbf{Cost of the worst-case guarantee.}]
The announced factor holds against the adversary, including a completion that spreads the probability of a heavy cluster
	across its whole skeleton.
A real model, however, concentrates that probability on a few items, often on ones that the first pass has already
	scored.
Thus, the realized gap is usually far below the announced factor, and a deployment reports both.
In other words, the announcement is pessimistic, whereas the planned run is not.
Apart from the speculative nodes of line~\ref{ln:spec} of \cref{alg:budget}, the planned run feeds only what every exact
	decoder must feed.

The speculative nodes are therefore the only cost of the announcement.
When the schedule forces them into a full batch, they replace prefixes that the real model may need, and since under the
	real model most of them turn out to hold no item of \ModelK, the overlap of the planned run with \ModelK can end a
	few percent below that of the unplanned run with the same budget.
When the announced factor is far above one, the useful output is the announcement itself, since it tells the deployment
	after one call that its budget is inadequate.
Moreover, \cref{eq:rstar} tells the deployment by how much, since for the given memory $M$ it gives the number of passes
	that exactness requires, and for the given number of passes $Y$ it gives the memory that exactness requires.
\end{remark}

\begin{remark}[\textbf{Deployment parameters.}]
\bmethod{} has two budgets and one target.
The budgets are $M$ and $Y$, and for $Y = L$ and $M = W$ they are the resources of beam search of width $W$.
Everything else in \bmethod{} follows from the budgets, the first pass, and the catalog trie, so there is no parameter
	to tune.
The target $\delta_t$ states what quality is sufficient, and it can only shorten the planned run: line~\ref{ln:stop} of
	\cref{alg:budget} returns as soon as \SidecarK is certified within $1 + \delta_t$, and with $\delta_t = 0$ it
	returns at exactness.
Note that this stopping rule rests on the certificate alone and needs no planning.

What the planning adds is that the deployment knows after the first call whether the target will be met within the
	budget, which is the case exactly when $\delta_t \ge \delta^*(M, Y)$, and how many passes exactness would take.
Hence, the deployment can serve, escalate, or commit to a latency before the fallback runs.
The budget and the target, however, do not play symmetric roles, since $Y$ binds unconditionally, whereas the target
	only allows the planned run to stop early.
Hence, the guarantee is $\max(\delta^*(M, Y), \delta_t)$: if $\delta_t \ge \delta^*$, the budget can deliver the target
	and the planned run may end early, and if $\delta_t < \delta^*$, it uses all $Y$ passes and \SidecarK is certified
	within $1 + \delta^*$, which the deployment learns after the first pass.
Finally, the draft is an input rather than a parameter to tune.
It determines how small $\delta^*$ is for given $M$ and $Y$, but it never affects whether \SidecarK and its certificate
	are correct.
\end{remark}

%% file: sections/05_experiments.tex
\section{Experiments}
\label{sec:exp}

We now evaluate \method{} and \bmethod{} against constrained beam search.
First, we describe the dataset, the model, and the replay procedure that allows us to validate our claims on every
	request.
Then, we introduce the baseline and the draft sources.
Finally, we state the research questions of interest to a practitioner and answer them in the rest of the section.

\paragraph{Data and model.}
For the purpose of this paper, we use the Beauty category of Amazon Reviews 2014~\citep{mcauley2015image,he2016ups},
	which is commonly used to evaluate SID-based generative retrieval approaches.
The dataset has 22,363 users and 12,101 items, and we split it with the leave-one-out protocol.
The SIDs come from an RQ-VAE with three codebooks of 256 codes each, followed by a collision code, so that every SID has
	$L = 4$ codes.
We train a four-layer decoder-only transformer on the histories of the users, written as sequences of SID codes.
It should be noted that \method{} only modifies the decoder stage, and therefore we do not focus on optimizing the
	hyperparameters of the transformer to obtain the most performant model.
However, the ranking quality of our trained model is within the confidence intervals of the published works using this
	setup.

\paragraph{Evaluation protocol.}
For every test user, we run one tree pass over all 21,091 internal nodes of the catalog trie to obtain the exact
	distribution of the model's probability mass.
Every decoder is then a simple deterministic replay of these stored distributions.
Thus, we can test every claim on every request, and compare the decoders on identical scores.
We use \ModelK with $K = 10$ as the reference target, since it is the output of exhaustive scoring of the model.
The goal of \method{} is to return the output of exhaustive scoring; no decoder can do better than the model that it
	decodes.
Thus, every comparison in this section uses \ModelK as the reference for both decoders.

\paragraph{Flatter models.}
In order to vary the sharpness of the model without changing the decoders, we apply a temperature $T$ to every
	conditional distribution of the trained model.
The trained model corresponds to $T = 1$, and a larger $T$ gives a flatter model.
Note that a temperature has a different effect here than on a single softmax that is commonly used in large language
	models.
For a single softmax, a temperature rescales the distribution but does not change the order of its outcomes.
Here, however, each of the $L$ conditionals is tempered at its own prefix and renormalized over the children of that
	prefix, so the ranking of the items changes.
As $T$ grows, the probability of an item tends to the product of one over the number of children of each prefix on its
	path.
The flatter models therefore rank the items differently from the trained model and serve as a useful proxy for a model
	trained on the noisy data that is common in large-scale recommender systems, since the conditionals of such a model
	are similarly spread out.

\paragraph{Baseline.}
Our baseline is constrained beam search with the same model and the same history cache, i.e., the setup used by TIGER.
We do not compare against the speculative decoders for SIDs, because each of them requires an artifact beyond the
	trained model.
AtSpeed~\citep{lin2025atspeed} requires a trained drafter, NEZHA~\citep{wang2026nezha} requires a trained self-drafting
	head, and SpecGR~\citep{ding2026specgr} requires a tuned acceptance threshold.
Moreover, these decoders aim to reproduce the list of beam search in fewer passes, and none of them certifies it.
Hence, \BeamK is also the reference against which these decoders are measured.

\paragraph{Potential draft sources.}
At request time, \method{} queries drafts from key-value stores.
Such drafts come predominantly from two kinds of sources, namely model-free and model-based sources.
The \emph{model-free} sources do not use the SID model at all.
Content $k$NN returns the items whose content embeddings are closest to any of the last five items of the history.
Co-occurrence and matrix factorization (implicit ALS) are built from the interaction log.
The \emph{model-based} sources, in contrast, derive the draft from the SID model, and the main one is the \emph{user
	sequence}, i.e., the top-$N$ list of the model for the user, which a batch job computes and stores.
Note that, throughout this section, the user sequence does not denote the items in the history of the user.
Instead, it denotes the $N$ items that the SID model itself ranks highest when a batch job runs it on the stale history
	of the user.
We consider three variants of the user sequence.
The first variant is the sequence stored by a copy of the model whose weights are quantized to FP4, which a cheaper
	batch job could run using constrained beam search.
The other two variants are copies of the sequence that are two and five interactions old.
Irrespective of the source, the drafts are always older than the request that triggered decoding.
RQ2 compares all seven sources, and from RQ4 onwards, in the interest of brevity, every figure uses two drafts of 500
	entries each.
The first draft is the user sequence, which is our default.
The second draft comes from matrix factorization, the strongest model-free source, which we henceforth refer to as the
	MF draft.

\paragraph{What the experiments measure.}
We checked every property proven in \cref{sec:onepass,sec:trie,sec:budgets} on every request of every experiment to
	ensure that none was violated.
Specifically, no certified list differed from \ModelK, no announced factor was exceeded, and no run took more passes
	than its worst-case bound.
Since correctness is thus settled, the following experiments focus on analyzing costs.
They address seven research questions (RQs), the first four with unlimited memory per pass and the other three under the
	budgets of beam search:
\begin{enumerate}[label=\textup{RQ\arabic*.}, leftmargin=2.6em, itemsep=0pt]
\item What decides the number of passes?
\item What is a good source of drafts?
\item How does the size $K$ of the output list affect the costs?
\item What incremental benefit does the catalog trie provide?
\item At beam search's resources, is \SidecarK at least as good as \BeamK?
\item What does exactness of \SidecarK cost?
\item What benefit does the worst-case guarantee of \bmethod{} provide, and what does it cost?
\end{enumerate}
Finally, we present the overall landscape of the results on one budget plane.

\paragraph{Reporting in the figures.}
Every figure in this section reports a statistic over requests, with one request per user.
The text that discusses each figure provides the context of its experiment, and wherever a setting is not mentioned, the
	figure uses 10,000 held-out users, drafts of $N = 500$ entries, and $K = 10$.
The error bars in the figures show 95\% confidence intervals over these requests, and they are often so narrow that they
	are hardly visible.

\begin{figure}[t]
\centering
\includegraphics[width=\columnwidth]{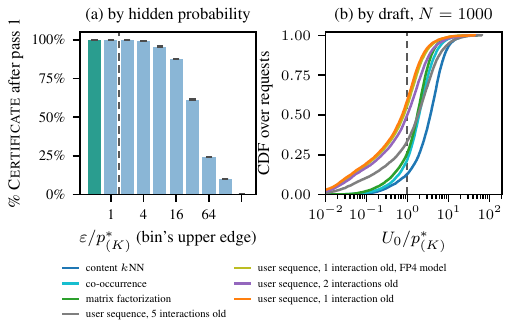}
\caption{\textbf{The hidden-probability condition.}
\textbf{(a)} Left of the line, the first pass is certified on every request, as proven, and it still is on most requests
	up to eight times the threshold.
\textbf{(b)} The exact condition, $U_0 < p^*_{(K)}$, orders the sources as their passes do.}
\label{fig:certificate}
\end{figure}

\begin{figure}[t]
\centering
\includegraphics[width=\columnwidth]{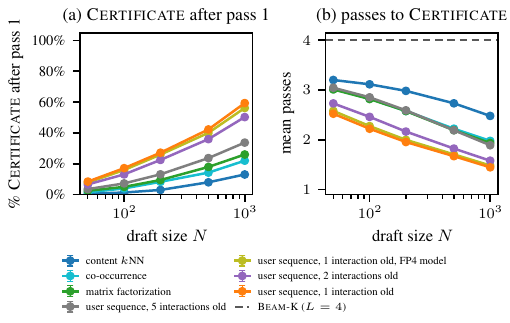}
\caption{\textbf{The draft sources.}
Every source yields an exact list, and the passes follow its alignment with the model: the model's own stale ranking
	needs well under two passes, even from an FP4 copy of the model, the behavioral sources about 2.2, and content
	similarity close to three.
Beam search always needs four passes.}
\label{fig:passes}
\end{figure}

\subsection{Number of Passes (RQ1)}
\label{sec:exp-components}

\begin{figure}[t]
\centering
\includegraphics[width=\columnwidth]{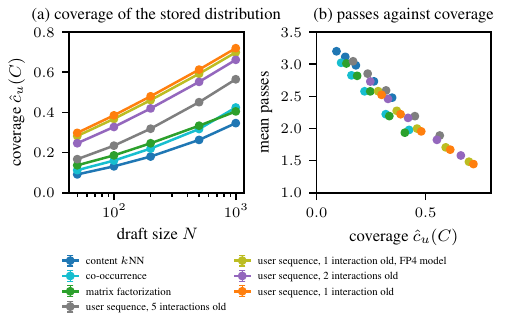}
\caption{\textbf{Alignment of the draft sources, measured offline.}
\textbf{(a)} The offline coverage $\hat c_u(C) = \sum_{e \in C} \hat p_u(e)$ of each draft against its size.
\textbf{(b)} The passes of \method{} against the offline coverage, one point per source and size.
All sources fall on one curve, so the offline coverage of a source predicts its number of passes and can be used to
	choose among sources.}
\label{fig:alignment}
\end{figure}

\begin{figure*}[tp]
\centering
\includegraphics[width=0.82\textwidth]{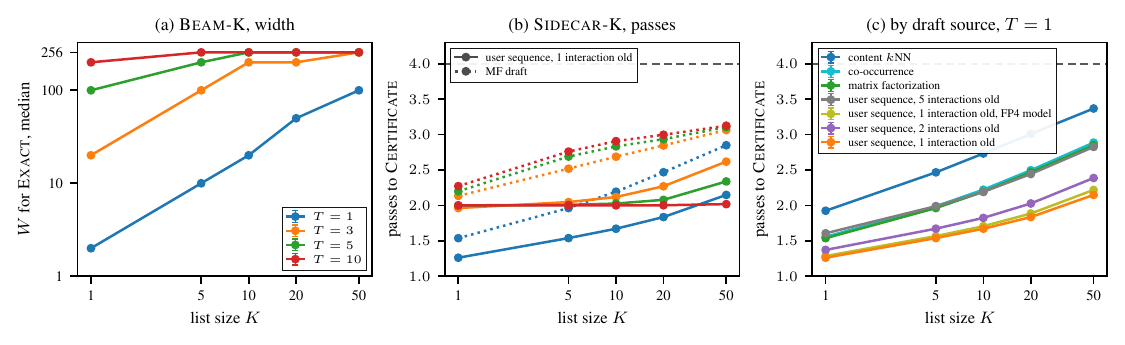}
\caption{\textbf{The size of the list.}
\textbf{(a)} The beam width for an exact list grows with $K$ on the trained model and sits at the largest width run on
	the flatter ones.
\textbf{(b)} \method{}'s passes rise gently with $K$ on the trained model, and with the user sequence they stay at two
	on the flat models whatever $K$ is.
\textbf{(c)} The order of the draft sources holds at every $K$.}
\label{fig:ksweep}
\end{figure*}

In this section, we address RQ1: what decides the number of passes?
Intuitively, the number of passes should depend on how much of the probability of the model the draft leaves uncovered.
The reason is that, by \cref{sec:onepass}, the first pass is certified whenever this hidden probability is below the
	probability of the $K$-th best item.

In order to test this intuition, we group the requests by the ratio of the hidden probability to the probability of the
	$K$-th best item.
We use unlimited memory per pass, so that the probability mass of the draft is the only factor that matters.
In \cref{fig:certificate}(a), we observe that the first pass is certified on every request to the left of the threshold,
	as proven.
Interestingly, the certificate still holds on most requests whose ratio is up to 8$\times$ the threshold.
This happens because the hidden probability is rarely concentrated in one place.
Instead, it is spread over many uncovered prefixes, and each of these prefixes is lighter than the $K$-th best item.
Hence, the sufficient condition of \cref{sec:onepass} is far from necessary.

Panel (b) shows the exactness condition, which compares the heaviest frontier prefix $U_0$ with the $K$-th best item, as
	a distribution over requests for each source at $N = 1000$.
The first pass is certified exactly on the requests to the left of one.
The sources whose drafts leave less probability uncovered place more of their requests to the left of this line: the
	curve of the user sequence lies furthest to the left, and the curve of content $k$NN furthest to the right.
These observations indicate that whenever the draft misses a cluster, the fallback needs at least one more round,
	because every level-1 prefix that the draft leaves uncovered has to be either opened or ruled out.
Thus, we conclude that the probability that the draft leaves uncovered, and in particular its heaviest uncovered prefix,
	decides the number of passes.

\subsection{Draft Sources (RQ2)}
\label{sec:exp-sources}

Next, we address RQ2: what is a good source of drafts?
Since every source yields \ModelK, the choice of source only affects the number of passes.
We find that the sources that are aligned with the SID model need the fewest passes, and that this alignment can be
	measured offline, before deployment.

\Cref{fig:passes} compares the seven sources with unlimited memory per pass.
Broadly, their costs fall into three separate tiers.
Content $k$NN needs close to three passes at any draft size, and its first pass is certified on fewer than 10\% of
	requests at 500 entries.
The two behavioral sources, which read the interaction log, need about 2.2 passes at 500 entries, and matrix
	factorization is slightly better than co-occurrence.
A user sequence that is one interaction old needs 1.7 passes, and its first pass is certified on about 40\% of requests.
The sequence stored by the FP4-quantized copy of the model needs only three hundredths of a pass more, although the
	weights of that copy are 11\% off and its top-10 agrees with the top-10 of the full model on only 69\% of requests.
A user sequence that is two interactions old, in contrast, needs another sixth of a pass.
A user sequence that is five interactions old needs 2.2 passes, which is the same as the MF draft, although it still
	leaves less probability uncovered and is certified after one pass more often.
Thus, \method{} tolerates both a cheaper quantized model and a stale history.

One may expect recall against the held-out item to explain this order.
However, the MF draft has nearly the same recall as the user sequence, yet it needs half a pass more.
Instead, the alignment of the draft with the model explains the order.
The behavioral sources contain items from the right clusters more often than content similarity does, which is why they
	need half a pass less than content $k$NN.
However, they rarely contain the top items of the model itself.

Moreover, this alignment can be measured before deployment.
Let $\hat p_u$ denote the distribution of the model for user $u$ one interaction ago.
The user sequence is the top-$N$ of this distribution, so the batch job that produces the user sequence has already
	computed $\hat p_u$.
We define the \emph{offline coverage} of a draft $C$ as the stored probability of its items, i.e., $\hat c_u(C) =
	\sum_{e \in C} \hat p_u(e)$.
A batch job can read this quantity off the stored table without running the decoder.
The offline coverage is the counterpart of $1 - \varepsilon$ for the model based on stale user history, computed over
	the items of the draft alone.
In \cref{fig:alignment}(a), we observe that the offline coverage ranks the sources exactly as their numbers of passes
	do.
Panel (b) plots the mean number of passes against the offline coverage, and every source at every size lies on a single
	curve.
This confirms the conclusion of RQ1 with a quantity that can be computed offline.
Thus, at the time of deployment, one can determine the best source and the size of the draft by simply choosing the
	source whose offline coverage is highest at the size permitted by the memory budget of the deployment.

\subsection{Size of the List (RQ3)}
\label{sec:exp-K}

\begin{figure*}[tp]
\centering
\includegraphics[width=0.82\textwidth]{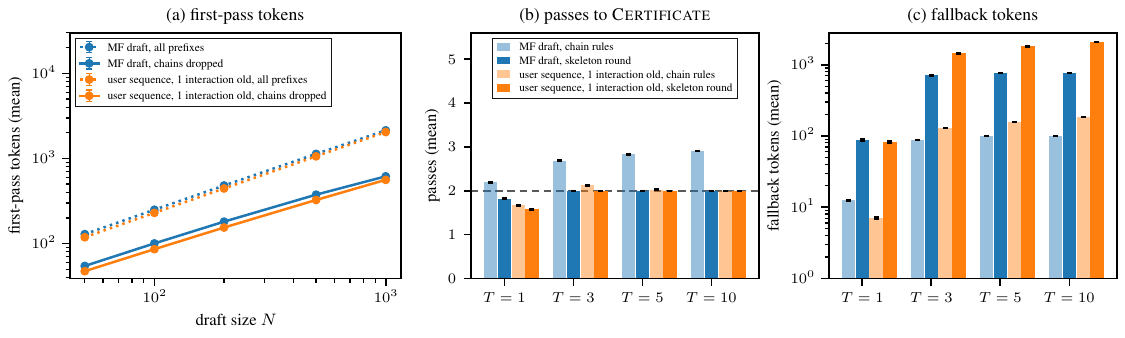}
\caption{\textbf{What the catalog trie saves.}
\textbf{(a)} Dropping the chain prefixes cuts the first pass to a third of its tokens.
\textbf{(b, c)} The skeleton round never needs a third pass, but it saves up to one pass only for the MF draft, and it
	feeds several times as many fallback tokens.}
\label{fig:skeleton}
\end{figure*}

Next, we address RQ3: how does the size $K$ of the output list affect the costs?
Like the sharpness of the model, the size of the list should be determined by the application rather than by the
	decoder.
We expect a longer list to make exactness harder, since the $K$-th best item is lighter and more prefixes can
	potentially outweigh it.
In order to quantify this effect, \cref{fig:ksweep} varies $K$ from $1$ to $50$ with unlimited memory per pass.

As expected, both decoders need more resources for a longer list, but they need different resources.
Beam search needs a wider beam (panel a).
On the trained model, the median request needs a width of 2 for the single best item and a width of 100 for the top-50.
On the flatter models, the median request needs the largest width that we run from $K = 5$ on.
\method{}, in contrast, needs more passes, and only a few more (panel b).
With the user sequence on the trained model, the number of passes rises from 1.2 to 2.2 across the same range of $K$.
On the flatter models, it stays at two for every $K$, because the first pass already holds the list and one fallback
	round certifies it, whatever its length.
Panel (c) shows that the order of the sources found in RQ2 holds at every list size, with the MF draft about half a pass
	behind the user sequence.
Thus, the size of the list affects the width of beam search by up to 50$\times$, whereas it only changes the number of
	passes of \method{} by about one and still requires fewer passes than beam search for $L = 4$.

\subsection{Benefit of the Catalog Trie (RQ4)}
\label{sec:exp-trie}

\begin{figure*}[tp]
\centering
\includegraphics[width=0.88\textwidth]{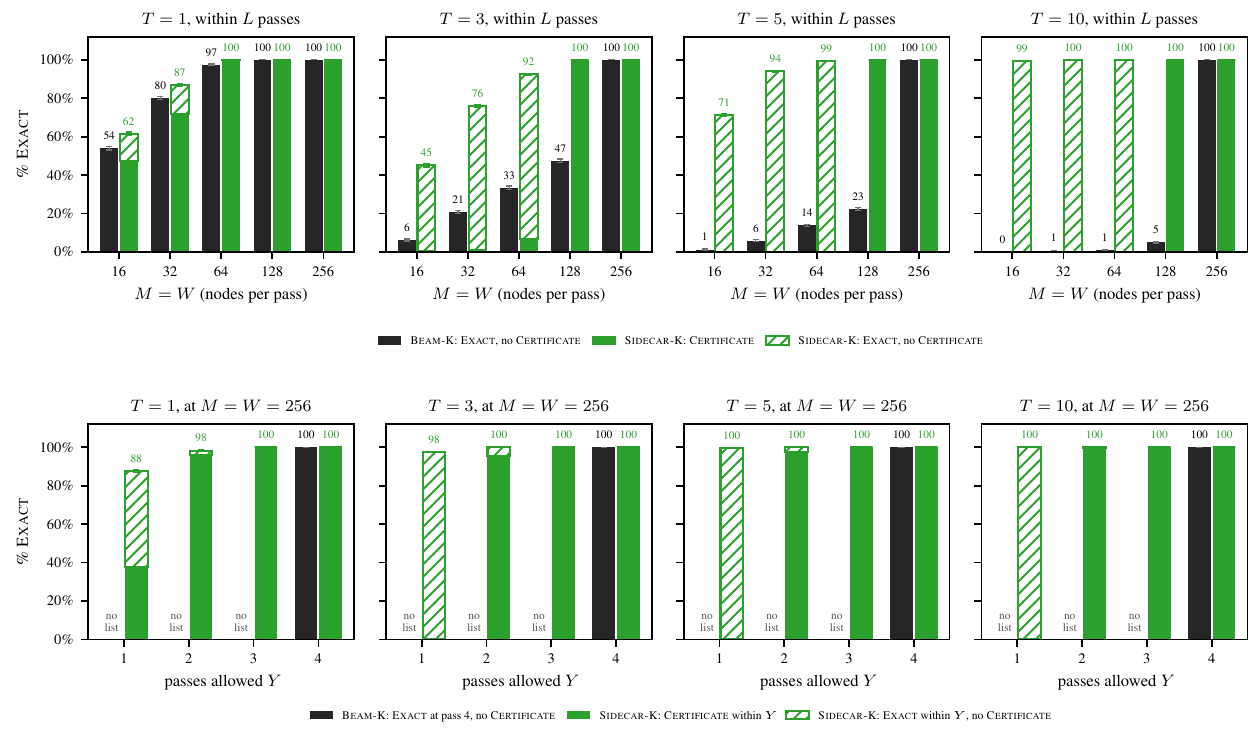}
\caption{\textbf{Equal resources against beam search.}
\textbf{Top:} beam search's latency of four passes, against memory.
\textbf{Bottom:} beam search's memory of 256 nodes, against passes.
On the sharp model the lists agree and only one decoder knows it.
On the flat models \SidecarK is exact at a fraction of the memory and two passes earlier.}
\label{fig:equal}
\end{figure*}

In this section, we address RQ4: what incremental benefit does the catalog trie provide?
In short, the structure of the catalog trie allows us to allocate tokens, i.e., memory, efficiently in the first pass,
	which can reduce the number of fallback passes required.
Two properties of the catalog trie make this possible.
First, the content codes address 16.7 million cells for only 12,101 items.
As a result, 78\% of the depth-2 prefixes and 98\% of the depth-3 prefixes hold a single item, and two thirds of the
	candidate prefixes of every draft lie on chains.
Second, the skeleton of the catalog consists of 2,533 of the 21,091 internal nodes of the trie, i.e., of about ten nodes
	per level-1 cluster.

\Cref{fig:skeleton} quantifies the effects of these two properties.
In panel (a), we observe that the chain rules of \cref{sec:trie} cut the first pass to a third of its tokens at every
	size, while the scores stay identical.
Panels (b) and (c) compare the two ways of resolving an open cluster, namely level by level and with its whole skeleton.
The skeleton round never needs a third pass if enough memory is available.
It reduces the passes of the MF draft from 2.2 to 1.8 on the trained model and from 2.9 to 2.0 on the flattest model.
However, the skeleton round feeds about 7$\times$ as many fallback tokens to achieve this reduction.
Moreover, it saves the user sequence little, because the chain rules alone already keep that draft near two passes.
Thus, feeding a skeleton whole saves a pass only when the draft leaves heavy clusters open, but it comes at the cost of
	additional tokens in fallback rounds.

\subsection{Equal Resource Comparison (RQ5)}
\label{sec:exp-equal}

\begin{figure*}[tp]
\centering
\includegraphics[width=0.71\textwidth]{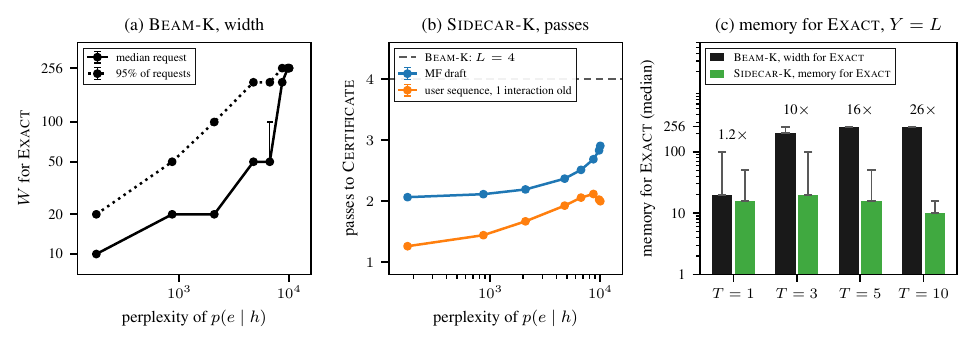}
\caption{\textbf{What exactness costs as the model flattens.}
Beam search needs a beam width that it cannot determine in advance (a), whereas \method{} needs a bounded number of
	passes (b).
On the memory axis (c) the gap opens to more than an order of magnitude on the flat models.}
\label{fig:cost}
\end{figure*}

Building upon these results with unlimited memory, we now compare the two decoders at equal resources and address RQ5:
	is \SidecarK at least as good as \BeamK?
In \cref{fig:equal}, beam search of width $W$ runs with its own resources, namely four passes with $W$ prefixes each.
\bmethod{} receives the same resources, namely $M = W$ nodes per pass and at most four passes, and all its passes,
	including the first pass, are capped at $M$ nodes.
We find that \SidecarK is at least as good as \BeamK at every budget, and much better where the model is flat.
Moreover, at every budget, \method{} states how close its list is to \ModelK, either with a certificate or with the
	announced factor, whereas beam search gives no indication of how far its list is from the exact one.

\paragraph{At beam search's latency.}
The top row of \cref{fig:equal} fixes the number of passes at four and varies the memory per pass.
On the trained model, the black and the green bars rise together and reach every request at the same memory.
Thus, \BeamK and \SidecarK coincide, and a deployment that looked only at the lists would see no difference.
The difference, however, lies in the solid part of the green bars, i.e., in the share of requests on which \method{} has
	proven its list exact.
From a modest memory on, this share covers nearly all requests.
Beam search, in contrast, has no such share at any width, because nothing in its computation distinguishes an exact list
	from a wrong one.
On the flat models, correctness itself separates the two decoders.
\SidecarK is exact on most requests at the smallest memory and on every request at a fraction of beam search's width,
	whereas \BeamK needs the full width to become exact.
It should be noted that at small memories, \method{} returns the exact list before it can prove that the list is exact,
	as the hatched part of the green bars shows.
The reason is that the open clusters cannot all be closed within the three fallback rounds.

\paragraph{At beam search's memory.}
The bottom row of \cref{fig:equal} fixes the memory per pass at beam search's width and varies the number of passes.
After a single pass, \method{} already holds \ModelK on most requests of the sharp model and on nearly all requests of
	the flat models.
At that point, beam search does not yet hold any list, since it needs three more passes to output a list.
One pass later, \method{} has certified nearly all of its lists.
Beam search has a single bar, at its fourth pass, where its lists are exact on these requests at this width.
Hence, a deployment that serves the list of \method{} after two passes compares favorably against one that waits for the
	four passes of beam search.

\paragraph{Why flatness helps \SidecarK and hurts beam search.}
It may seem counterintuitive that the performance of \bmethod{} improves with the flatness of the underlying model.
Here, we explain the reasoning behind it.
This pattern follows from what determines the top-$K$ of a flat model.
The probability of an item is a product of $L$ conditionals.
As the model flattens, each conditional tends to one over the number of admissible children, so the probability of an
	item tends to the product of these reciprocal branching factors along its path.
The top-$K$ of a flat model then consists of the items in small subtrees, which share the probability of their cluster
	with few siblings, regardless of the history.
Indeed, on the trained model, the clusters that hold the top-10 are among the heaviest clusters of the trie.
On the flattest model, in contrast, these clusters have median probability mass, and all of them are below the median
	size.
This mechanism has three manifestations.
First, the user sequence of a flat model, which is its ranking for the stale history, almost surely contains the top
	items for the current history, because the ranking is now determined by the shape of the catalog rather than by the
	history.
Thus, staleness matters less and less as the model flattens, and the first pass already returns the exact list.
Second, beam search ranks prefixes by their probability mass.
It therefore keeps the heavy clusters and drops the light ones, although the light clusters are the ones that hold the
	top items.
This is why the bars of beam search collapse on the flat models until its width covers every cluster.
Third, the certificate has to rule out the heavy clusters.
Each frontier prefix of a heavy cluster outweighs the small top items, even though none of these prefixes holds one of
	them, and ruling these prefixes out takes rounds.
In other words, flatness separates where the probability mass is from where the best items are located in the trie.

\subsection{Cost of Exactness (RQ6)}
\label{sec:exp-cost}

\begin{figure*}[tp]
\centering
\includegraphics[width=0.82\textwidth]{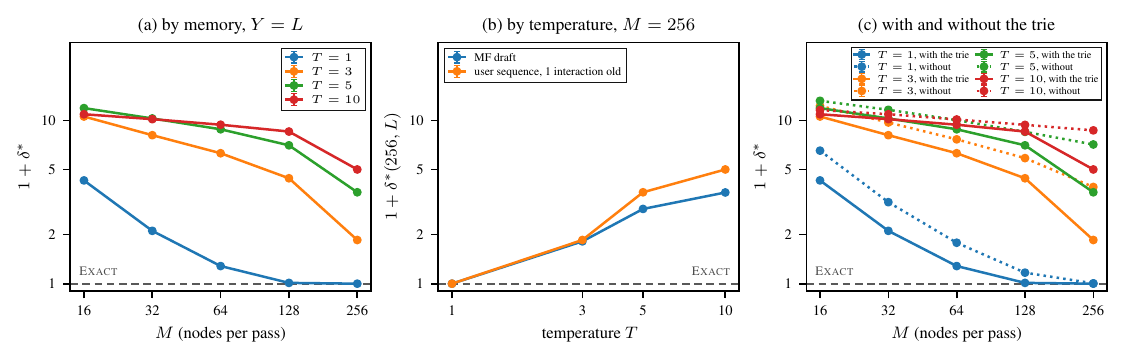}
\caption{\textbf{The announced factor}, against memory at four passes (a) and against temperature at beam search's
	memory (b).
It rises as the model flattens for both drafts, and at beam search's memory it is lower for the MF draft, whose
	candidates spread over more clusters, although that draft is exact far less often at smaller memories.
The realized gap was zero at every budget.
\textbf{(c)} With the catalog trie (solid) the factor is never looser than the count-based announcement without it
	(dotted, \cref{eq:nhat}), and it is about half of it on the flat models at beam search's width.}
\label{fig:factor}
\end{figure*}

\begin{figure}[t]
\centering
\includegraphics[width=\columnwidth]{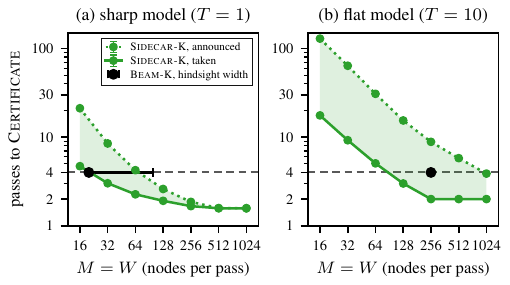}
\caption{\textbf{Passes against memory for a certified list.}
Beam search is a point at its four passes, and \bmethod{} is a curve below and to its left on both models.}
\label{fig:frontier}
\end{figure}

Next, we address RQ6: what does exactness of \SidecarK cost?
\Cref{fig:cost} varies the sharpness of the model and reports the resources required by each decoder to output an exact
	list.
We find that on a sharp model the scarce resource is the number of passes, whereas on a flat model it is the memory per
	pass.
On both kinds of model, \method{} needs less of the resource that is scarce.

Beam search needs width, i.e., memory for a wider beam (panel a).
A flat model spreads its probability over many prefixes of nearly equal mass, so the beam has to be wide enough to keep
	the right ones.
Across the sweep of temperatures, the width that the median request needs grows more than 10$\times$.
From a moderate flatness on, the width that 95\% of requests need reaches the largest width that we test.
Moreover, the width required to achieve exactness for beam search is a property of the request that beam search learns
	only in hindsight, so a deployment has to provision every request assuming the worst-case scenario.
\method{}, in contrast, needs extra passes to achieve exactness (panel b), and this cost is bounded by construction.
With the user sequence, it levels off at two passes on the flat models, and with the MF draft, it rises toward three.

Panel (c) compares the two decoders on the memory axis, since memory per pass determines how many requests a serving
	machine can process at once, and hence the size of the serving fleet.
On the trained model, both decoders need about the same memory for an exact list.
There, \method{} gains in passes, since at equal memory it needs fewer than half of beam search's passes.
As the model flattens, however, the memory that beam search needs grows, whereas the memory that \method{} needs does
	not.
The gap opens to more than an order of magnitude, and at the flat end it is 10--26$\times$.
This is a large difference in favor of \bmethod{} that affects the practicality of the two decoders.

\Cref{fig:frontier} shows both resources in one plane.
Beam search is a single point in the plane of memory and passes, at four passes and the width that the median request
	needs.
\bmethod{}, in contrast, is a curve, since more memory per pass reduces the number of passes to a certified list.
On both models, the curve lies below and to the left of the point.
Hence, both a deployment that is short of memory and a deployment that is short of latency gain from \method{}.

\subsection{Cost vs.\ Benefit of the Worst-Case Guarantee (RQ7)}
\label{sec:exp-guarantee}
\label{sec:exp-price}

\begin{figure}[t]
\centering
\includegraphics[width=\columnwidth]{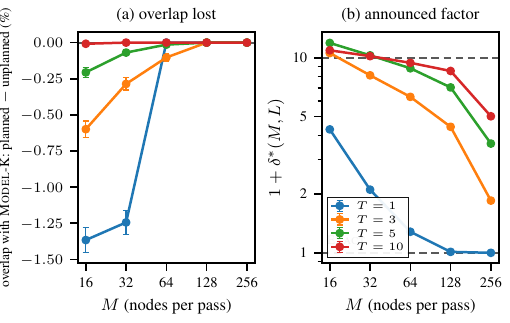}
\caption{\textbf{The price of the announcement}, user sequence.
\textbf{(a)} The overlap of \bmethod{}'s list with \ModelK minus that of the unplanned run, in percent.
\textbf{(b)} The announced factor at the same budgets.
The price is at most one or two percent, and only at the smallest memories.}
\label{fig:price}
\end{figure}

\begin{figure*}[tp]
\centering
\includegraphics[width=0.92\textwidth]{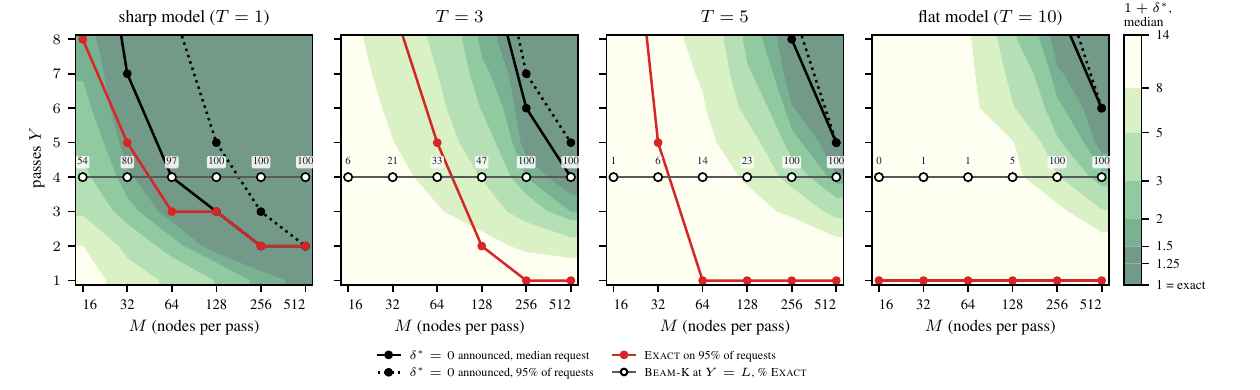}
\caption{\textbf{The budget map.}
Color: the factor announced after the first pass at each budget, darkest for exactness.
Black: the smallest $Y$ at which exactness is announced, for the median request (solid) and for 95\% of requests
	(dotted).
Red: the smallest $Y$ at which \SidecarK is exact on 95\% of requests.
Beam search is the row $Y = L$ with its share of exact lists.
The announced and the realized frontier coincide on the sharp model and part ways on the flat one.}
\label{fig:budgetmap}
\end{figure*}

Having measured what the planned run achieves, we finally address RQ7: what benefit does the worst-case guarantee of
	\bmethod{} provide, and what does it cost?
The guarantee is the announced factor $1 + \delta^*$ of \cref{thm:plan}, which holds within the budget against the
	adversary described in \cref{sec:budgets}.
An unplanned run has no such guarantee: until its last pass, it cannot say how far its list is from \ModelK.
In the worst case, it returns no item of \ModelK at all, even though the same budget suffices for exactness when the run
	is planned (\cref{prop:separation}).
We find that the announced factor reaches exactness within beam search's budget on the trained model, and that the
	guarantee lowers the overlap with \ModelK only at budgets where the factor is too large to be useful.

\paragraph{Size of the announced factor.}
\Cref{fig:factor} shows the announced factor in three panels.
In panel (a), the number of passes is fixed at four, as for beam search, and the memory per pass varies.
On the trained model, the factor drops to one at beam search's width.
A deployment with that budget therefore knows after the first pass that its list will be \ModelK within four passes.
On the flat models, the factor falls more slowly with memory, because every open cluster carries more probability and
	the adversary may hide a top item anywhere below it.

In panel (b), the memory per pass is fixed at beam search's width and the temperature varies.
The factor rises as the model flattens.
It is lower for the MF draft than for the user sequence, because a behavioral draft spreads its 500 candidates over more
	clusters and leaves less probability uncovered.
Exactness, however, ranks the two drafts the other way, as shown in RQ5.
Thus, the announcement measures how much probability a draft leaves uncovered, whereas exactness depends on whether the
	draft contains the top items.
For the user sequence, the realized gap was zero at every one of these budgets, so only the certification fell short.

Panel (c) compares the announcement of \cref{thm:plan} with an announcement that uses only the number of items below
	each prefix (\cref{prop:notrie} in Appendix~\ref{app:further}).
Including the catalog trie never loosens the factor, and it roughly halves it on the flat models at beam search's width.
The reason is that, without the trie, the announcement must assume that every open cluster is resolved level by level,
	which can take a node for every item below it.
With the trie, a cluster can instead be resolved by feeding its whole skeleton in one round, which takes only about ten
	nodes.
This is a further benefit of the trie, in addition to the chain rules and skeleton rounds of RQ4.

\paragraph{Overlap lost to the guarantee.}
To guarantee the worst case, \bmethod{} uses some of the nodes in each fallback pass for speculative prefixes, which the
	real model may not need.
The unplanned run uses all of its nodes for prefixes that are known to be open, so its list can overlap more with
	\ModelK.
\Cref{fig:price}(a) shows this difference.
For the user sequence, the two runs overlap equally from a moderate memory on, and \bmethod{} loses 1--2\% of overlap
	below that memory.
The loss occurs only at small memories, where the announced factor is large (\cref{fig:price}(b)).
Even there, the announcement tells the deployment after the first pass that its memory is too small, and \cref{eq:rstar}
	gives the number of fallback rounds that it would need to certify its list.
It should be noted that whether speculation helps or hurts the overlap depends on the model as much as on the procedure.
The speculative nodes are the nodes that the adversary would need.
Hence, they improve the list when heavy clusters hide items of \ModelK, and they slightly lower the overlap when the
	heavy clusters hold no such items.
On this benchmark, the top items of a flat model lie in light clusters (RQ5), which is the second case.
Hence, a deployment should announce the factor, which is the commitment that it can make before the fallback runs, and
	report the realized gap afterwards, which can only be smaller.

\subsection{Overall Landscape}
\label{sec:exp-recap}

\Cref{fig:budgetmap} summarizes the section on a single visual plane.
Every point of the plane is a budget of $M$ nodes per pass and $Y$ passes.
The color at a point is the factor that \bmethod{} announces at that budget after its first pass.
Thus, for any budget set by a deployment, the map shows what \bmethod{} can promise before the fallback runs.
The map shows two frontiers.
The dashed curve marks where \SidecarK is exact on nearly every request, as measured by RQ5 and RQ6.
The solid curve marks where this exactness is also announced after one call, which is the guarantee of RQ7.
Beam search is the row $Y = L$, and it announces nothing at any budget.

On the sharp model, the two frontiers nearly coincide, and both cross the row of beam search at a fraction of its
	required width.
Thus, at budgets where \BeamK is merely exact, \SidecarK is exact, announced as exact, and obtained in a fraction of the
	passes.
On the flat model, the two frontiers diverge.
The realized frontier drops to a single pass at every memory shown, whereas the announced frontier retreats to the top
	right corner.
As a result, at beam search's own budget, the guarantee is in the form of the announced factor rather than exactness.
Between the two frontiers, \SidecarK is exact, but the announcement cannot yet say so.
This region is wide on the flat model because the real model concentrates probability where the draft already looks,
	whereas the announcement must allow for the worst case, in which the adversary spreads that probability elsewhere.

The map thus separates the two decisions that a practitioner has to make at the time of deployment.
Above the dashed curve, the deployment can expect an exact list.
Above the solid curve, it can also commit to exactness after one call.
Between the two curves, it receives \SidecarK together with a factor that it can state.
In practical terms, when the model is sharp, \bmethod{} is preferable outright, since it returns the list of \BeamK in a
	fraction of the passes and certifies it.
When the model is flat, memory is the deciding resource, and \SidecarK is exact with a fraction of the memory of beam
	search or in half of its passes.
When neither resource suffices for exactness, \bmethod{} still states how close \SidecarK is to \ModelK.

\paragraph{Early stopping.}
The certificate and the announced factor also allow a deployment to stop early, in three ways that beam search cannot
	offer.
First, the decoder stops as soon as its list is certified.
The bottom row of \cref{fig:equal} shows that, at beam search's memory, nearly all lists are certified after two passes,
	so the remaining passes of the budget are not spent.
Second, a deployment that accepts an approximate list can set a target factor and stop as soon as the list is certified
	within it.
The colors of \cref{fig:budgetmap} show, for every budget, the factor that is announced after the first call on the
	median request, and hence the budgets at which a given target is met.
Third, a deployment with a latency deadline can stop after any pass.
The bottom row of \cref{fig:equal} shows that the list after a single pass is already exact on most requests of the
	sharp model and on nearly all requests of the flat models, and the dotted curve of \cref{fig:frontier} shows after
	the first call how many passes the certificate will need.
In all three cases, the deployment can also act on the announcement itself: when the announced factor after the first
	call exceeds its target, it can stop at once and serve the list with its stated factor, or escalate the request to a
	larger budget.

%% file: sections/06_related_work.tex
\section{Related work}
\label{sec:related}

To the best of our knowledge, \method{} is the first decoder for Semantic IDs that proves its list to be the exact
	top-$K$ of the model or, under a budget, states a factor for its list before the fallback runs.
Our work connects to several streams of research, which we discuss below.

\paragraph{Generative retrieval and Semantic IDs.}
One line of research frames retrieval as the generation of document identifiers: the Differentiable Search
	Index~\citep{tay2022dsi} introduced the idea, and the Neural Corpus Indexer~\citep{wang2022nci} and query-generation
	variants~\citep{zhuang2022dsiqg} developed it for documents.
Subsequent work studied its limits at corpus scale~\citep{pradeep2023scale}, its reading as dense retrieval over
	prefixes~\citep{nguyen2023grdense}, and ranking objectives over identifiers~\citep{li2024ltrgr,zeng2024ripor}.
TIGER~\citep{rajput2023tiger} brought Semantic IDs to recommendation with codes from an RQ-VAE~\citep{lee2022rqvae}, in
	the manner of residual quantization in neural codecs~\citep{zeghidour2021soundstream}, and decoded them with
	constrained beam search.
Item identifiers for language-model recommenders were studied by \citet{geng2022p5,hua2023indexing}.
Another line of research learns the tokenizer from collaborative
	signals~\citep{jin2024lmindexer,zheng2024lcrec,wang2024letter,zhu2024cost,liu2025etegrec,qu2024tokenrec,
	si2024seater,wang2024eager}.
Moreover, Semantic IDs were shown to improve generalization in ranking models~\citep{singh2024semanticids}.
The depth of the SID tree is still an open choice: an extra level can improve recommendation up to a
	point~\citep{fang2026prism}, a shallower SID has improved offline and online metrics~\citep{xie2026singlelevel}, and
	a deeper TIGER degraded under beam search~\citep{hou2025rpg}, although depth changes the serving cost and the model
	at the same time.
Industrial systems, such as PLUM~\citep{he2025plum}, GLIDE~\citep{damico2026glide},
	OneRec~\citep{deng2025onerec,onerec2025report}, COBRA~\citep{yang2025cobra}, and the generative recommenders of
	Meta~\citep{zhai2024actions,yang2024liger}, all decode with beam search over the SID trie, and a practitioner's
	study catalogs their design choices~\citep{ju2025handbook}.
Although these works improve what the model ranks, they are orthogonal to ours, as they leave the decoding of the
	ranking unchanged.
In comparison, we leave the model and its tokenizer unchanged and change only the decoder.
Our approach applies to any of these tokenizers.

\paragraph{Beam search and its alternatives.}
\citet{oh2026personalized} showed that a single head with beam search gives better ranking and content quality than
	multi-token prediction~\citep{gloeckle2024mtp}, which parallel SID decoders~\citep{hou2025rpg, zhan2026stamp} adopt
	for throughput.
\citet{wu2025constrained} showed that constrained beam search can reach high top-1 precision while recalling branches
	poorly at top-$k$.
Before Semantic IDs, tree-based deep models~\citep{zhu2018tdm, zhu2019jtm} and Deep
	Retrieval~\citep{gao2021deepretrieval} retrieved with beam search over learned node scores, in the lineage of
	hierarchical softmax~\citep{morin2005hierarchical}, and \citet{zhuo2020beamtree} showed that beam search over such
	trees is not Bayes-optimal and trained the scorers to compensate.
These works adapt the model to the decoder; in contrast, we take the opposite route, in which the model stays as trained
	and the decoder becomes exact and certified.

\paragraph{Constrained decoding and exact search.}
Constrained decoding limits each next token to the choices that are allowed.
GENRE~\citep{decao2021genre} did so with a trie of entity names, and PICARD and grammar-constrained decoding did so with
	a grammar~\citep{scholak2021picard,geng2023gcd}.
Decoding Semantic IDs is the simplest such case, because every item is a path of the same length in a trie, and the
	probability of a prefix is the total probability of the items below it.
Hence, a prefix is never more likely than its own parent; exact search methods rely on this fact.
These methods include exact search for the single best sequence~\citep{stahlberg2019nmt} and best-first beam
	search~\citep{meister2020bestfirst,meister2020beam}.
However, they search for one sequence of unbounded length, where exactness is expensive.
In contrast, we search for the top-$K$ list over a trie of fixed depth $L$, where one pass over a good draft often
	suffices to certify the list, and $L$ passes always do.
A similar idea appears in classical retrieval, where WAND and block-max WAND~\citep{broder2003wand,ding2011bmw} certify
	a top-$K$ list using score upper bounds that are computed offline.
In contrast, our upper bounds come from the same forward pass that scores the draft, and the fallback expands only the
	prefixes that violate them.

\paragraph{Speculative decoding.}
Speculative decoding~\citep{leviathan2023speculative,chen2023speculative,sun2023spectr} drafts tokens cheaply and checks
	them against the target in one pass.
Tree attention extends this to a whole draft tree~\citep{miao2024specinfer, cai2024medusa}, and subsequent work added
	draft heads on the target~\citep{ankner2024hydra,li2024eagle}, lookahead and
	self-speculation~\citep{fu2024lookahead,zhang2024draftverify}, staged and big-little
	drafters~\citep{spector2023staged, kim2023biglittle}, retrieval-based drafts~\citep{he2024rest}, and draft trees
	chosen under a token budget~\citep{chen2024sequoia}.
Perhaps closest to our objective are the speculative decoders for generative recommendation:
	AtSpeed~\citep{lin2025atspeed} adapts speculative decoding to beam search and trains an aligned drafter,
	NEZHA~\citep{wang2026nezha} adds a self-drafting head, SpecGR~\citep{ding2026specgr} drafts by retrieval and accepts
	by a likelihood threshold, and STARK~\citep{unisgr2026} shares the KV cache across beams.
These decoders accelerate generation, but none of them certifies its list as we do.
In contrast, we use the existing draft without training, score the whole draft in one pass, and either certify the
	resulting output list or correct it with the fallback.

\paragraph{Search under budgets.}
Our certificate (\cref{thm:cert}) has the form of the stopping rule of the threshold algorithm for top-$K$
	aggregation~\citep{fagin2003optimal}, i.e., the $K$-th best score seen against a bound on everything unseen.
Moreover, the request-dependent cost set by the hidden probability (\cref{sec:onepass}) is of the kind that Fagin's
	instance optimality describes.
Similarly, any decoder that certifies its list must score every prefix whose probability exceeds the $K$-th best item's
	score, which parallels the nodes that best-first search must expand under any admissible
	heuristic~\citep{dechter1985generalized}.
The additive depth term in the budgeted round bound has the shape of the lower bound for exploring a tree whose
	structure is not known in advance~\citep{fraigniaud2006collective,dereniowski2015fast}, which no exploration rule
	can remove.
The catalog trie removes that uncertainty, because the decoder knows in advance which prefixes
	exist~\citep{papadimitriou1991shortest}.
Hence, \cref{thm:plan} quantifies how much knowing the trie helps.
Finally, the factor that we announce is the same kind of guarantee as in anytime and contract
	search~\citep{likhachev2003ara,aine2010contract}, which return a solution within a stated factor of the best one;
	our plan certifies that factor rather than estimating it.
Sequoia~\citep{chen2024sequoia} also chooses its speculation tree under a token budget, but by expected acceptance,
	whereas our plans are worst-case.

\paragraph{Serving.}
Our layout relies on paged KV caching~\citep{kwon2023pagedattention}, prefix sharing~\citep{zheng2024sglang}, continuous
	batching~\citep{yu2022orca}, and fused and block-sparse attention
	kernels~\citep{dao2022flashattention,dong2024flexattention}, which are orthogonal to the decoder and apply to
	\method{} as they do to beam search.

%% file: sections/07_impact_statement.tex
\section*{Impact Statement}
This paper presents a decoding procedure for generative recommenders that returns the same list as the model's own
	ranking at a lower cost in sequential model calls, together with a certificate that states whether the list is
	exact, or how close to exact it is.
It does not change what is recommended, but only how fast and how reliably the model's ranking is produced.
Lower decoding cost makes it cheaper to serve such recommenders, and a certificate makes their behavior under a budget
	auditable, which we view as positive.
The societal consequences of recommendation systems themselves, including their effects on exposure and diversity, are
	the responsibility of the model and of the stack that deploys it, and are not altered by the decoder.

%% file: sections/appendix_A_assumptions_and_proofs.tex
\section{Assumptions and proofs}
\label{app:proofs}

In this appendix, we state the assumptions under which the results of the paper hold, and we prove the results in the
	order in which they appear in the main text.
Each assumption is introduced in the main text where it is first needed.

\paragraph{Assumptions.}
Throughout the appendix, we assume the following.
\begin{enumerate}[label=\textbf{A\arabic*}, leftmargin=2.6em, itemsep=1pt]
\item Every SID has exactly $L$ codes, and the catalog trie is prefix-closed.
\item The scores are constrained to the admissible children of a prefix and sum to one over them; hence, the probability
	of a prefix equals the total probability of the items below it.
\item The output of the tree pass at a node equals the output of a separate pass over the path of that node.
\item There are no ties among the probabilities that the statements compare.
\item At least $K$ items are scored after the first pass, so that $s_K(0) > 0$.
\item Unless a budget is stated, a fallback round expands every open prefix.
\end{enumerate}

\paragraph{The frontier.}
Every proof below uses the following property of the frontier, which we state and prove first.

\begin{lemma}[frontier]
\label{lem:frontier}
After any round, every item either has an exact score or lies below exactly one frontier prefix, and in the second case
	its probability is at most that of the frontier prefix.
\end{lemma}

\begin{proof}
Fix an item $e$ and follow its SID from the root, which counts as expanded, down to the first prefix of $e$ that is not
	expanded.
Two cases arise.
\begin{enumerate}[label=(\alph*), itemsep=1pt]
\item If every prefix of $e$ up to depth $L-1$ is expanded, then the depth-$(L-1)$ prefix of $e$ is expanded, and hence
	$e$ has an exact score.
\item Otherwise, the first prefix of $e$ that is not expanded has an expanded parent, so it is a frontier prefix.
By A2, the probability of $e$ is at most the probability of this prefix.
\end{enumerate}
Moreover, no frontier prefix is an ancestor of another, since a frontier prefix is not expanded, whereas the parent of
	every frontier prefix is.
Hence, the frontier prefixes cover disjoint sets of items, and the frontier prefix found in case (b) is unique.
\end{proof}

\begin{proof}[Proof of \cref{thm:cert}]
Let $e$ be an item outside $E_r$.
By \cref{lem:frontier}, $e$ lies below some frontier prefix $b$, and therefore
\[
p(e) \;\le\; p(b) \;\le\; U_r \;\le\; s_K(r),
\]
where the first inequality follows from \cref{lem:frontier}, the second from the definition of $U_r$ as the largest
	frontier probability, and the third from the assumption of the theorem.
Hence, every item outside $E_r$ has a probability of at most the $K$-th best score in $E_r$, and since A4 excludes ties,
	it ranks below the $K$ best items of $E_r$.
Thus, the top-$K$ of $E_r$ is the top-$K$ of the catalog, i.e., $\SidecarKr{r} = \ModelK$.
\end{proof}

\begin{proof}[Proof of \cref{prop:anydraft}]
The proposition makes three claims, which we prove in turn.
\begin{enumerate}[label=(\alph*), itemsep=1pt]
\item \emph{$s_K$ never decreases.}
A round only adds scored items, so $E_{r+1} \supseteq E_r$, and the $K$-th best score in a larger set cannot be smaller.
\item \emph{$U$ never increases.}
A round removes every expanded prefix $b$ from the frontier and adds the children of $b$.
By A2, the probabilities of these children sum to $p(b)$, so each of them is at most $p(b)$.
Hence, the largest frontier probability cannot rise.
\item \emph{At most $L$ passes.}
First, a frontier prefix that is closed after some round stays closed in every later round, since its probability does
	not change and, by (a), $s_K$ never decreases.
Second, by A6, round $r+1$ expands every prefix that is open after round $r$.
Hence, every prefix that is open after round $r+1$ is a child of a prefix that was open after round $r$, and the
	smallest depth among the open prefixes grows by at least one per round.
Since frontier prefixes have depths between $1$ and $L-1$, no prefix is open after $L-1$ fallback rounds, i.e., after at
	most $L$ passes.
At that point $U \le s_K$, and \cref{thm:cert} applies.
\end{enumerate}
\end{proof}

\begin{proof}[Proof of \cref{prop:gap}]
We prove the three inequalities of the proposition from left to right.
\begin{enumerate}[label=(\alph*), itemsep=1pt]
\item \emph{$s_k(r) \le p^*_{(k)}$.}
Since $E_r$ is a subset of the catalog, its $k$-th best score cannot exceed the $k$-th best probability in the catalog.
\item \emph{$p^*_{(k)} \le \max(s_k(r), U_r)$.}
We distinguish two cases.
If $p^*_{(k)} > U_r$, then each of the $k$ best items of the catalog has a probability above $U_r$.
By \cref{lem:frontier}, an item without an exact score has a probability of at most $U_r$, so all $k$ of these items are
	scored, and together with (a) we obtain $s_k(r) = p^*_{(k)}$.
Otherwise, $p^*_{(k)} \le U_r$ holds directly.
In both cases, $p^*_{(k)} \le \max(s_k(r), U_r)$.
\item \emph{$\max(s_k(r), U_r) \le (1 + \max(\gap_r, 0))\, s_k(r)$.}
The first term satisfies $s_k(r) \le (1 + \max(\gap_r, 0))\, s_k(r)$ trivially.
For the second term, the definition of the gap gives $U_r = (1 + \gap_r)\, s_K(r)$, and since $k \le K$ implies $s_k(r)
	\ge s_K(r)$, we obtain $U_r \le (1 + \max(\gap_r, 0))\, s_k(r)$.
\end{enumerate}
\end{proof}

\input{sections/fig_proof_plan}

\begin{proof}[Proof of \cref{prop:plancost}]
\Cref{fig:proof-plan} illustrates both parts of the proof.

(i) Two bounds hold for the nodes at level $j$ below $f$ that are above $t$ and have to be fed.
\begin{enumerate}[label=(\alph*), itemsep=1pt]
\item \emph{The probability bound.}
The prefixes at one level below $f$ are disjoint and, by A2, have total probability $p(f)$.
Since each prefix above $t$ contributes more than $t$ to this total, at most $\floor{p(f)/t} = k_f$ of them exceed $t$.
\item \emph{The skeleton bound.}
Every node outside the skeleton of $f$ lies on a chain, i.e., its subtree holds a single item, so by A2 its probability
	equals the probability of that item.
The output at the parent of the top node of the chain already gives this probability, so such a node needs no token,
	whether or not it is above $t$.
Hence, the nodes that have to be fed at level $j$ are among the $c_j(f)$ skeleton nodes at that level.
\end{enumerate}
Together, the two bounds give at most $\min(c_j(f), k_f)$ such nodes at level $j$.

(ii) Fix any consistent completion (\cref{sec:budgets}).
Under this completion, the nodes above $t$ form a set that is closed under taking ancestors, since by A2 a node has at
	least the probability of each of its descendants.
The plan $\ell$ resolves $f$ in two phases.
\begin{enumerate}[label=(\alph*), itemsep=1pt]
\item \emph{Descent.}
For the levels $j = 0, \dots, \ell - 1$, one level per round, the plan feeds the open nodes of level $j$, of which there
	are at most $\min(c_j(f), k_f)$ by (i).
This gives the first term of \cref{eq:plan}.
\item \emph{Skeleton round.}
The plan then feeds the open nodes of level $\ell$, of which there are at most $k_f$, together with their skeletons,
	which include the nodes themselves.
The total size of these skeletons is at most the sum of the $k_f$ largest skeleton sizes at level $\ell$, i.e.,
	$\Sigma^\sigma_\ell(f, t)$, which is the second term of \cref{eq:plan}.
\end{enumerate}
After both phases, every node below $f$ has either been fed or lies below a closed node.
A closed node bounds every item in its subtree by $t$, so nothing below $f$ remains open.
\end{proof}

\input{sections/fig_proof_rounds}

\begin{proof}[Proof of \cref{prop:rounds}]
(i) We classify every fallback round by the number of pool nodes that it feeds: a round is \emph{full} when it feeds $M$
	pool nodes, and \emph{under-filled} otherwise (\cref{fig:proof-rounds}).
We call a pool node \emph{available} once the output at its parent is known, and we bound the two kinds of rounds
	separately.
\begin{enumerate}[label=(\alph*), itemsep=1pt]
\item \emph{Full rounds.}
Every pool node is fed at most once, and the pool holds at most $S$ nodes, so at most $\ceil{S/M}$ rounds are full.
\item \emph{Under-filled rounds.}
The procedure never leaves an available pool node unfed while a slot is free, so an under-filled round feeds every
	available pool node.
After such a round, every remaining pool node lies at least one level deeper than before, so the largest number of
	levels that a remaining plan still descends one round at a time drops by one.
Hence, at most $m$ under-filled rounds occur while such levels remain.
\item \emph{The last round.}
Once no such level remains, every remaining pool node is available, so every round is full except possibly the last one,
	which empties the pool.
If this last round feeds $a \ge 1$ nodes, the full rounds number at most $\floor{(S - a)/M} \le \ceil{S/M} - 1$, so the
	last round is already counted in $\ceil{S/M}$.
\end{enumerate}
The two counts add up regardless of the order in which full and under-filled rounds occur, which gives the bound
	$\ceil{S/M} + m$.

(ii) Let $B = \ceil{S/M} + m$, suppose that $B \le y$, and let primes denote the quantities after the round.
The proof has three steps.
\begin{enumerate}[label=(\alph*), itemsep=1pt]
\item \emph{The schedule fits into one pass.}
Since $\ceil{S/M} \le y - m$, we have $y - 1 - m \ge \ceil{S/M} - 1$, and therefore
	\[
	x_r \;=\; S - (y - 1 - m)\, M \;\le\; S - \bigl(\ceil{S/M} - 1\bigr)\, M \;\le\; M.
	\]
\item \emph{The bound holds with one round fewer.}
Let $x$ denote the number of pool nodes that the round feeds, and assume for now that the pool does not grow and that no
	plan becomes deeper, i.e., $S' \le S - x$ and $m' \le m$; step (c) justifies this.
We distinguish two cases.
\begin{itemize}[itemsep=1pt] \item If $x \ge x_r$, then $S' \le S - x_r = (y - 1 - m)\, M$, so $\ceil{S'/M} \le y - 1 -
	m$, and together with $m' \le m$ we obtain $B' \le y - 1$.
\item If $x < x_r$, the round has fed every available pool node.
Every remaining node above $t$ is then a child of a node that was fed without its skeleton, i.e., of a node whose plan
	descends at least one more level.
Hence, $m' \le m - 1$ while $S' \le S$, and again $B' \le y - 1$.
If $m = 0$, every pool node is available from the start, so in this case the round has fed the whole pool, and the pool
	is empty afterwards.
\end{itemize}
\item \emph{Re-planning does not enlarge the pool.}
After the round, the pool and the plans are recomputed, and three facts ensure that this does not increase $S$ or $m$.
\begin{itemize}[itemsep=1pt] \item The threshold can only rise between rounds, since $s_K$ never decreases, and a higher
	threshold only shrinks the set of nodes above it, and hence the pool.
\item For every fed node $g$, the allotment of the children of $g$ never exceeds the allotment of $g$ minus the node
	fed.
Indeed, their counts $k_c$ satisfy $\sum_c k_c \le k_g$, because their probabilities sum to $p(g)$, a sum of floors is
	at most the floor of the sum, and the threshold has not fallen.
Moreover, the inequality $\min(a, b) + \min(a', b') \le \min(a + a', b + b')$ bounds their level counts by those of the
	parent, and their $\Sigma^\sigma$ terms select at most $k_g$ nodes at one level below $g$.
\item Re-choosing the plans after the round can only lower $B'$, since the previous plans remain available.
\end{itemize}
\end{enumerate}
\end{proof}

\begin{proof}[Proof of \cref{thm:plan}]
(i) Let $\Delta$ denote the announced factor, and let $t = (1 + \Delta)\, s_K(r)$ denote the corresponding threshold
	after round $r$.
\begin{enumerate}[label=(\alph*), itemsep=1pt]
\item By the definition of $\delta^*$ in \cref{eq:deltaplan}, after the first pass some plans satisfy $\ceil{S/M} + m
	\le y$, where $y = Y - 1$ is the number of remaining rounds.
\item Every round of \bmethod{} feeds at least $x_r$ pool nodes or every available one (lines~\ref{ln:batch}
	and~\ref{ln:spec} of \cref{alg:budget}).
Hence, by \cref{prop:rounds}(ii), the bound holds again with one round fewer after every round, i.e., it is an invariant
	of the planned run.
\item When no round remains, the bound reads $\ceil{S/M} + m \le 0$, which forces $S = 0$, i.e., no frontier node is
	above $t$.
Hence, after the last round $R$ we have $U_R \le (1 + \Delta)\, s_K(R)$, and \cref{prop:gap} gives (i).
\end{enumerate}

(ii) With $Y = Y_{\min}(M)$ passes, the announced factor is $\delta^*(M, Y) = 0$, so by (i) with $\Delta = 0$, no
	frontier node is open after $Y_{\min}(M) - 1$ rounds.
Hence, \cref{thm:cert} applies, and \SidecarK is certified.

(iii) Apply \cref{prop:rounds}(ii) with the factor announced after round $r$.
It shows that after the round, some plans satisfy $\ceil{S'/M} + m' \le y - 1$ at the new threshold.
Since $\delta^*_{r+1}$ is the smallest factor for which such plans exist with one round fewer, it is at most
	$\delta^*_r$.
\end{proof}

\input{sections/fig_proof_sep}

\begin{proof}[Proof of \cref{prop:separation}]
We construct the instance for any $L \ge 3$, $K \ge 1$, and $b \ge 2$ with $b^{L-1} \ge K$, with $M = (b^{L-1} -
	1)/(b-1)$ nodes per pass and any $2 \le Y \le L - 1$ passes.
We then show that the unplanned run, i.e., \cref{alg:main} with $M$ nodes per pass, returns after $Y$ passes a list
	without any item of \ModelK and is exact only after $L$ passes, whereas \bmethod{} announces exactness after the
	first pass and returns \ModelK with its certificate after two passes.
The proof has four steps.

\emph{Step 1: the instance} (\cref{fig:proof-sep}).
\begin{itemize}[itemsep=1pt]
\item \emph{The trie.}
The root has two children, $A$ and $B$.
Below $A$, the trie is a complete $b$-ary tree: every prefix of depth $1$ to $L-1$ below $A$ has $b$ children, and the
	prefixes of depth $L$ are the items.
Hence, $A$ holds $b^{L-1}$ items, every prefix below $A$ is branching, and the skeleton of $A$ is the whole tree of
	internal nodes, $\sigma(A) = 1 + b + \dots + b^{L-2} = (b^{L-1} - 1)/(b-1) = M$.
Below $B$, there are $K$ children, each of which is a chain down to a single item, so $B$ holds $K$ items.
\item \emph{The model.}
The model puts probability $\eta$ on $B$, split evenly over its $K$ items, and probability $1 - \eta$ on $A$, split over
	its items with pairwise distinct probabilities that lie within a factor of two of each other.
For example, uniform conditionals that are perturbed slightly satisfy A2 and A4.
We choose $\eta$ small enough that every item below $A$ has a probability above $\eta/K$.
Then \ModelK consists of $K$ items below $A$, and since the conditionals below $A$ are nearly uniform, these items lie
	below distinct depth-$(L-1)$ prefixes.
\item \emph{The draft.}
The draft consists of the $K$ items below $B$.
\end{itemize}

\emph{Step 2: the first pass}, which is common to both runs.
The candidate trie is $B$ with its $K$ chain children, so the first pass with the chain rules feeds $B$ alone, and the
	chain rules resolve the $K$ items below $B$ exactly.
Hence, $E_0$ is the draft, $s_K(0) = \eta / K$, and the frontier is $F_0 = \{A\}$ with $p(A) = 1 - \eta > s_K(0)$, so
	$A$ is open and the certificate fails.

\emph{Step 3: the unplanned run.}
\begin{itemize}[itemsep=1pt]
\item Every prefix below $A$ is branching, so no chain rule applies, and every fallback round feeds the open prefixes
	one level deeper.
After $r$ rounds, the fed prefixes below $A$ are therefore those of depths $1, \dots, r$, with at most $b^{L-2} \le M$
	of them per round, so the budget never binds.
\item An item below $A$ is scored only when its depth-$(L-1)$ prefix is fed, which first happens after $L - 1$ rounds,
	i.e., after $L$ passes.
\item Hence, after $Y \le L - 1$ passes, the scored set is still the draft, so the returned list consists of the $K$
	items below $B$ and contains no item of \ModelK.
Moreover, the frontier holds prefixes of probability about $(1 - \eta)/b^{Y-1} > s_K$, so the certificate fails as well.
\item After $L$ passes, every item below $A$ is scored, the list equals \ModelK, and the frontier is empty.
\end{itemize}

\emph{Step 4: the planned run (\bmethod{}).}
\begin{itemize}[itemsep=1pt]
\item \emph{The announcement.}
At any threshold $t = (1 + \delta)\, s_K(0) < 1 - \eta$, the only open prefix is $A$, and its plan $\ell = 0$ costs
	$S_0(A, t) = \sigma(A) = M$ by \cref{eq:plan}.
Hence, $S^*_0(t) \le M$ and $R^*(t) \le \ceil{M/M} + 0 = 1 \le Y - 1$ for every $Y \ge 2$.
It follows that $\delta^*(M, Y) = 0$ and $Y_{\min}(M) = 2$, i.e., exactness is announced after the first pass.
\item \emph{The first fallback round.}
The pool is $A$ with its skeleton, so $S = M$ and $m = 0$, and the schedule demands $x_1 = S - (Y - 2)\, M \le M$ pool
	nodes.
The batch therefore holds $A$ followed by its $M - 1$ remaining skeleton nodes, which fit into the pass.
\item \emph{The result.}
After this round, every depth-$(L-1)$ prefix below $A$ has been fed, every item below $A$ is scored, and no frontier
	prefix remains.
Hence, $U_1 = 0$, the certificate holds, and $\SidecarKr{1} = \ModelK$.
Thus, the planned run returns \ModelK with its certificate after two passes, for every $Y \ge 2$.
\end{itemize}
\end{proof}

%% file: sections/fig_proof_plan.tex
\begin{figure}[htb]
\centering
\resizebox{0.62\textwidth}{!}{%
\begin{tikzpicture}[x=1cm, y=1cm, font=\footnotesize, every node/.style={inner sep=1.5pt}]
\coordinate (f) at (0,0);
\coordinate (c1) at (-1.9,-1.1); \coordinate (c2) at (0,-1.1); \coordinate (c3) at (1.9,-1.1);
\coordinate (d1) at (-2.5,-2.2); \coordinate (d2) at (-1.3,-2.2); \coordinate (d3) at (-0.5,-2.2); \coordinate (d4) at
(0.5,-2.2);
\coordinate (d5) at (1.4,-2.2); \coordinate (d6) at (2.4,-2.2);
\draw (f)--(c1) (f)--(c2) (c1)--(d1) (c1)--(d2) (c2)--(d3) (c2)--(d4);
\draw[black!45] (f)--(c3) (c3)--(d5) (c3)--(d6);
\foreach \n/\w in {d1/0.5, d2/0.35, d3/0.45, d4/0.35, d5/0.4, d6/0.3} {
  \path (\n) ++(0,-0.28) coordinate (t\n);
  \draw[black!45, fill=white] (t\n) -- ++(-\w,-0.85) -- ++(2*\w,0) -- cycle;
}
\draw[draw=blue!60!black, fill=blue!10, thick] (td1) -- ++(-0.5,-0.85) -- ++(1.0,0) -- cycle;
\draw[draw=blue!60!black, fill=blue!10, thick] (td3) -- ++(-0.45,-0.85) -- ++(0.9,0) -- cycle;
\node[pfopen] at (f) {$f$};
\node[pfopen] at (c1) {}; \node[pfopen] at (c2) {}; \node[pfclosed] at (c3) {};
\node[pffed] at (d1) {}; \node[pfclosed] at (d2) {}; \node[pffed] at (d3) {}; \node[pfclosed] at (d4) {};
\node[pfclosed] at (d5) {}; \node[pfclosed] at (d6) {};
\node[pflvl] at (-3.2,0) {level 0}; \node[pflvl] at (-3.2,-1.1) {level 1}; \node[pflvl] at (-3.2,-2.2) {level 2};
\node[pflvl] at (-3.2,-2.95) {skeletons};
\node[pfann] at (3.0,0) {round 1: $f$};
\node[pfann] at (3.0,-1.1) {round 2: open children,\\$\le \min(c_1, k_f) = 2$};
\node[pfann] at (3.0,-2.2) {round 3: open nodes,\\$\le k_f = 2$, with skeletons};
\node[pfann, text=black!55] at (-3.2,-3.75) {gray: closed, bounds its subtree};
\end{tikzpicture}}
\caption{\textbf{What a plan can cost} (\cref{prop:plancost}), for an open prefix $f$ with $k_f = \lfloor p(f)/t \rfloor
	= 2$ and the plan $\ell = 2$.
At most $k_f$ nodes per level can be above the threshold, and the open nodes (blue) form a top part of the tree, since
	an open node has an open parent.
A closed node (gray) bounds its whole subtree, so nothing below it is fed.
The plan feeds the top part one level per round down to level $\ell - 1$, at most $\min(c_j, k_f)$ nodes at level $j$,
	and then the open nodes of level $\ell$ with their whole skeletons (shaded triangles), which gives the two terms of
	\cref{eq:plan}.}
\label{fig:proof-plan}
\end{figure}
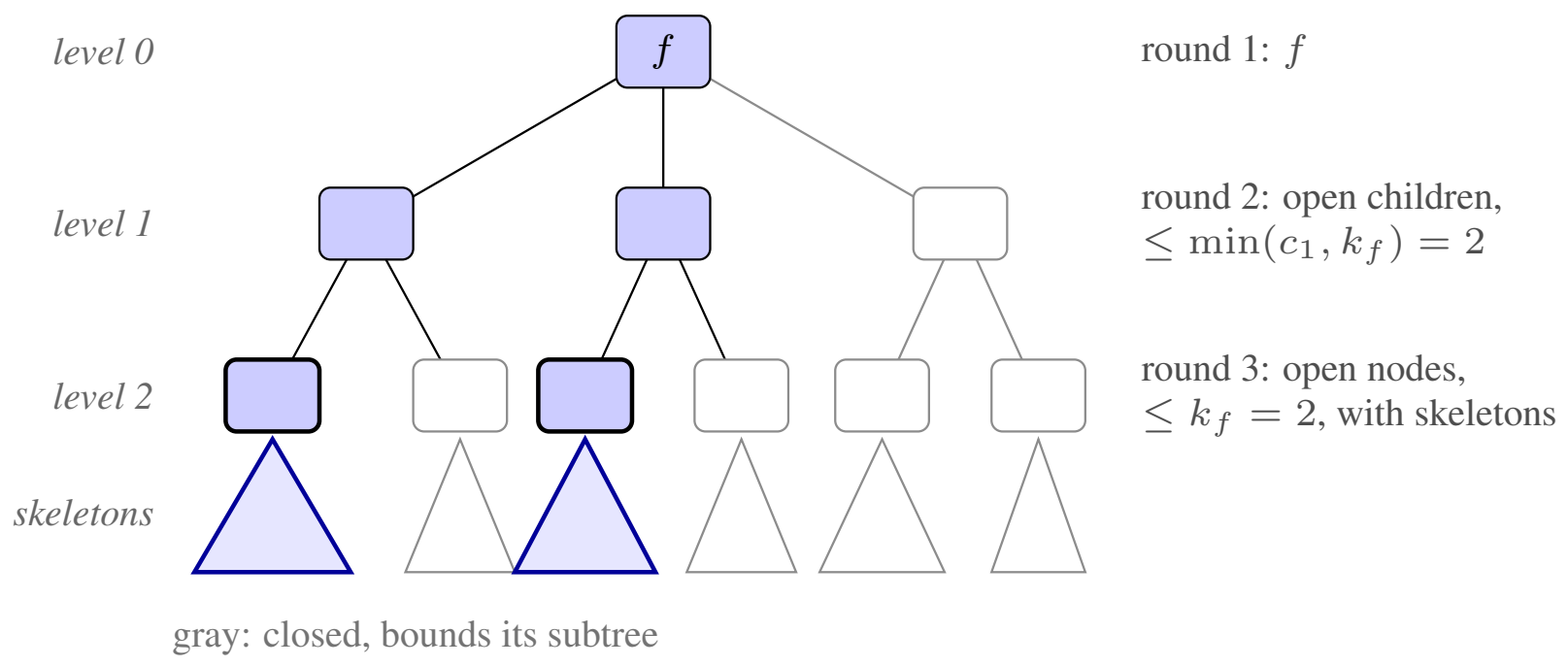

%% file: sections/fig_proof_rounds.tex
\begin{figure}[htb]
\centering
\resizebox{0.5\textwidth}{!}{%
\begin{tikzpicture}[x=1cm, y=1cm, font=\footnotesize, every node/.style={inner sep=1.5pt}]
\foreach \r/\k/\lab in
{1/6/full, 2/6/full, 3/3/under-filled{,} depth $2 \to 1$, 4/6/full, 5/2/under-filled{,} depth $1 \to 0$, 6/4/last{,}
empties the pool} {
  \pgfmathsetmacro{\y}{-0.62*\r+0.35}
  \node[pflvl] at (-0.1,\y) {round \r};
  \foreach \i in {1,...,6} {
    \pgfmathparse{\i<=\k ? 1 : 0}
    \ifnum\pgfmathresult=1 \node[pffull] at (0.48*\i-0.24,\y) {}; \else \node[pfempty] at (0.48*\i-0.24,\y) {}; \fi
  }
  \node[pfann] at (3.0,\y) {\lab};
}
\node[pfann] at (-1.2,-4.15) {full or last: $4 \le \lceil 27/6 \rceil = 5$\\under-filled: $2 \le m = 2$};
\end{tikzpicture}}
\caption{\textbf{The round bound} $\lceil S/M \rceil + m$ (\cref{prop:rounds}(i)), here with $M = 6$ slots per round, a
	pool of $S = 27$ nodes and depth $m = 2$.
Every round either fills its $M$ slots with pool nodes or runs out of available pool nodes.
Rounds of the first kind, together with the last round, number at most $\lceil S/M \rceil$, since every pool node is fed
	once.
A round of the second kind has fed every available pool node, so the remaining depth drops by one, which happens at most
	$m$ times.
The two counts add in whatever order the two kinds occur.}
\label{fig:proof-rounds}
\end{figure}
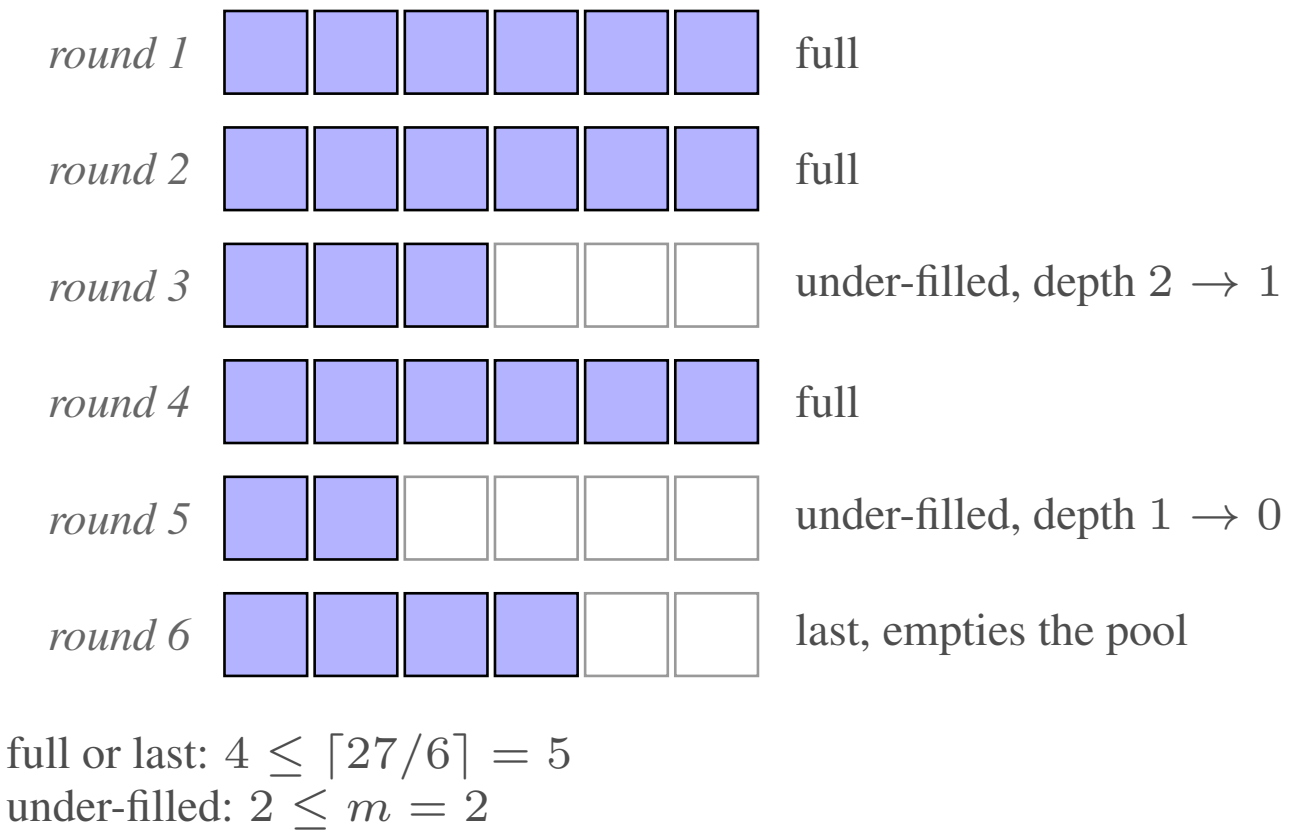

%% file: sections/fig_proof_sep.tex
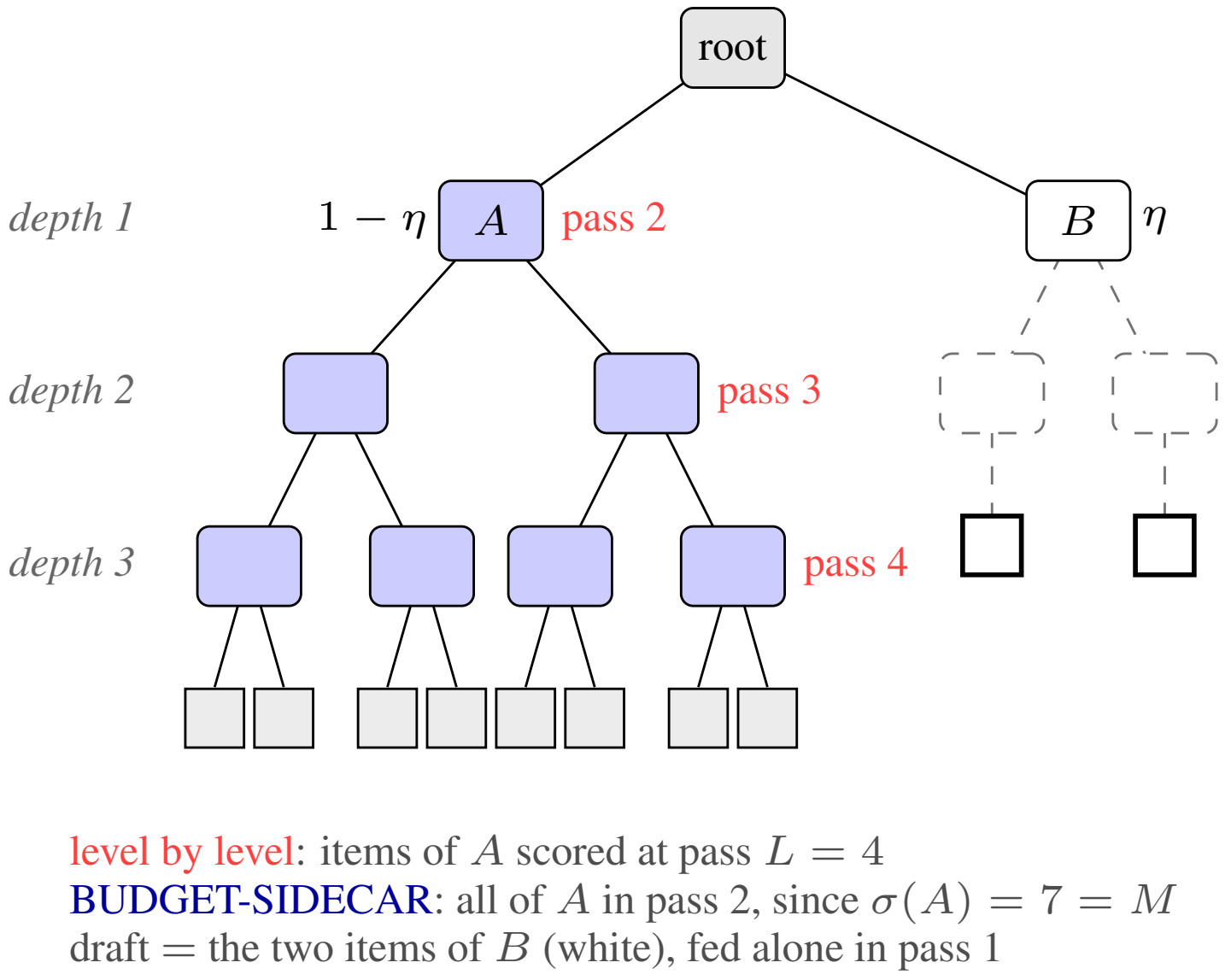
\begin{figure}[htb]
\centering
\resizebox{0.55\textwidth}{!}{%
\begin{tikzpicture}[x=1cm, y=1cm, font=\footnotesize, every node/.style={inner sep=1.5pt}]
\coordinate (root) at (2.6,0);
\coordinate (A) at (1.2,-1.0); \coordinate (B) at (4.6,-1.0);
\coordinate (a1) at (0.3,-2.0); \coordinate (a2) at (2.1,-2.0);
\coordinate (a11) at (-0.2,-3.0); \coordinate (a12) at (0.8,-3.0); \coordinate (a21) at (1.6,-3.0); \coordinate (a22) at
(2.6,-3.0);
\coordinate (b1) at (4.1,-2.0); \coordinate (b2) at (5.1,-2.0);
\draw (root)--(A) (root)--(B) (A)--(a1) (A)--(a2) (a1)--(a11) (a1)--(a12) (a2)--(a21) (a2)--(a22);
\draw[dashed, black!55] (B)--(b1) (B)--(b2) (b1)--++(0,-0.7) (b2)--++(0,-0.7);
\foreach \n in {a11,a12,a21,a22} {\draw (\n)--++(-0.2,-0.7) (\n)--++(0.2,-0.7);
  \node[pfitem] at ([shift={(-0.2,-0.88)}]\n) {}; \node[pfitem] at ([shift={(0.2,-0.88)}]\n) {};}
\node[pfitemc] at ([shift={(0,-0.88)}]b1) {}; \node[pfitemc] at ([shift={(0,-0.88)}]b2) {};
\node[pfnd, fill=gray!20] at (root) {root};
\node[pfopen, label={[font=\scriptsize]left:$1-\eta$}] at (A) {$A$};
\node[pfnd, fill=white, label={[font=\scriptsize]right:$\eta$}] at (B) {$B$};
\node[pfopen] at (a1) {}; \node[pfopen] at (a2) {};
\foreach \n in {a11,a12,a21,a22} {\node[pfopen] at (\n) {};}
\node[pfnd, dashed, draw=black!55, fill=white] at (b1) {}; \node[pfnd, dashed, draw=black!55, fill=white] at (b2) {};
\node[font=\scriptsize, text=red!75, anchor=west] at ([shift={(0.35,0)}]A) {pass 2};
\node[font=\scriptsize, text=red!75, anchor=west] at ([shift={(0.35,0)}]a2) {pass 3};
\node[font=\scriptsize, text=red!75, anchor=west] at ([shift={(0.35,0)}]a22) {pass 4};
\node[pflvl] at (-0.8,-1.0) {depth 1}; \node[pflvl] at (-0.8,-2.0) {depth 2}; \node[pflvl] at (-0.8,-3.0) {depth 3};
\node[pfann] at (-1.3,-4.95) {\textcolor{red!75}{level by level}: items of $A$ scored at pass $L = 4$\\
\textcolor{blue!60!black}{\bmethod{}}: all of $A$ in pass 2, since $\sigma(A) = 7 = M$\\
draft $=$ the two items of $B$ (white), fed alone in pass 1};
\end{tikzpicture}}
\caption{\textbf{The instance of \cref{prop:separation}}, with $b = 2$, $L = 4$, $K = 2$ and $M = 7$.
The heavy cluster $A$ holds probability $1 - \eta$ spread evenly over its items, and every prefix below it branches, so
	its skeleton is the whole tree of exactly $\sigma(A) = M$ nodes.
The light cluster $B$ holds $\eta$ on two chains, and its two items are the draft, which the first pass feeds alone,
	leaving $A$ open.
Level-by-level descent (red) reaches the items of $A$ only at pass $L = 4$, so with fewer passes it returns the items of
	$B$.
\bmethod{} finds $R^*(t) = \lceil 7/7 \rceil = 1$ at every threshold, announces exactness after the first pass, and
	feeds the skeleton of $A$ whole in the second.}
\label{fig:proof-sep}
\end{figure}

%% file: sections/appendix_B_further_guarantees.tex
\section{Further guarantees}
\label{app:further}

Building upon the results of \cref{sec:onepass,sec:trie,sec:budgets}, this appendix states further guarantees that
	sharpen them.
The main text does not use these guarantees.

\paragraph{Alignment by depth.}
Let $\mu_j$ be the probability of the items whose first $j$ codes match no candidate of the draft, so that $\mu_0 = 0
	\le \mu_1 \le \dots \le \mu_{L-1} = \varepsilon$.
These quantities measure how the draft is aligned with the model, depth by depth, and they refine the hidden-probability
	condition of \cref{sec:onepass}.

\begin{proposition}[alignment by depth]
\label{prop:alignment-more}
\begin{enumerate}[label=\textup{(\roman*)}]
\item The certificate holds after one pass if $U_0 < p^*_{(K)}$, and only if $U_0 \le p^*_{(K)}$.
\item After $r$ fallback rounds, \SidecarKr{r} misses at most $\floor{\mu_{L-1-r} / p^*_{(K)}}$ items of \ModelK.
\item Every fallback round expands at most $\floor{\varepsilon / s_K(0)}$ prefixes.
\end{enumerate}
\end{proposition}

Part (i) is the sharp form of that condition: the first pass certifies exactly when the heaviest uncovered prefix is
	lighter than the $K$-th best item, and \cref{fig:certificate} shows the condition on the benchmark.
Part (ii) says that each fallback round repairs one level of misalignment.
An item that the draft missed at depth $j$ is reached after $L - j$ rounds, and the probability left uncovered at that
	depth bounds how many such items can exist.
Part (iii) bounds the number of tokens that a round feeds.

\begin{proof}
(iii) Every open prefix after round $r$ has probability above $s_K(r) \ge s_K(0)$.
Open prefixes are disjoint, since they are frontier prefixes, and their total probability is at most $\varepsilon$,
	because every item below a frontier prefix is unscored after the first pass and stays unscored below an unexpanded
	prefix.
Hence, fewer than $\varepsilon / s_K(0)$ of them exist.

(ii) Let $e$ be an item of \ModelK that is missing from \SidecarKr{r}.
Then $e$ is unscored, so it lies below a frontier prefix $b$ of the first pass, of some depth $d$, and $p(b) \ge p(e)
	\ge p^*_{(K)} \ge s_K(r')$ for every round $r'$, with strict inequality while $e$ is unscored by A4.
So $b$ is open after the first pass and round one expands it (A6).
The depth-$(d+1)$ prefix of $e$ is then open and round two expands it, and so on, until the depth-$(L-1)$ prefix of $e$
	is expanded and $e$ is scored.
This takes $L - d$ rounds.
Since $e$ is still missing after $r$ rounds, $d \le L-1-r$, which means that the first $L-1-r$ codes of $e$ match no
	candidate.
The missing items are therefore distinct items of probability at least $p^*_{(K)}$ among a set of total probability
	$\mu_{L-1-r}$ (A2), so at most $\floor{\mu_{L-1-r}/p^*_{(K)}}$ of them exist.

(i) If $U_0 < p^*_{(K)}$, every item of the exact top-$K$ is scored, since an unscored one would lie below a frontier
	prefix of probability at least $p^*_{(K)}$.
Hence, $s_K(0) = p^*_{(K)} > U_0$ and the certificate holds.
Conversely, the certificate gives $U_0 \le s_K(0) \le p^*_{(K)}$.
\end{proof}

\paragraph{Exact top ranks of an early-stopped list.}
\Cref{prop:gap} bounds every rank of a list returned before the certificate holds.
In fact, the top ranks of such a list are already exact.

\begin{proposition}[exact top ranks]
\label{prop:head}
After any round $r$, if $s_k(r) \ge U_r$ for some $k \le K$, the first $k$ returned items are exactly the model's
	top-$k$.
\end{proposition}

\SidecarKr{r} therefore agrees with \ModelK down to the first rank whose score falls below $U_r$, and only the ranks
	below it are bounded by the factor of \cref{prop:gap} rather than exact.

\begin{proof}
If $s_k(r) \ge U_r$, every unscored item has probability at most $s_k(r)$, so the $k$ best scored items are the $k$ best
	of the catalog (A4).
\end{proof}

\paragraph{Announcement without the trie.}
It helps to see what the catalog trie contributes to the announcement of \cref{sec:budgets}.
Without it, the only information about the nodes below an open frontier prefix $f$ is the probability bound of
	\cref{prop:plancost}(i), at most $k_f = \floor{p(f)/t}$ open nodes at any level, and the number $n_f$ of items below
	$f$, which caps the nodes at any level as well.
The fallback can then only go level by level, and the open prefixes must fit in a single round at every level.
This gives the count
\beq{
\widehat N(t) = \sum_{f \in F_0:\, p(f) > t} \Bigl(1 + (L-1-\depth(f))\, \min\bigl(n_f, k_f\bigr)\Bigr)
\label{eq:nhat}
}
of prefixes that can ever be open, and hence the announcement $\delta^*(M) = \min\{\delta : \widehat N((1+\delta)\,
	s_K(0)) < M\}$ for $Y = L$ passes.
The trie improves on this in two places: $c_j(f)$ replaces $n_f$ as the structural cap, and the skeleton plans add a
	second way to resolve an open prefix within the budget, namely by feeding its whole skeleton in one round.
The following proposition shows that the trie never loosens the announcement.

\begin{proposition}[the announcement without the trie]
\label{prop:notrie}
\begin{enumerate}[label=\textup{(\roman*)}]
\item With $M$ nodes per pass and $Y = L$ passes, the level-by-level fallback returns a list within a factor $1 +
	\delta^*(M)$ of the exact top-$K$ at every rank.
\item $\delta^*(M, L) \le \delta^*(M)$: the announcement with the trie is never looser than the one without it.
\end{enumerate}
\end{proposition}

\Cref{fig:factor}(c) in \cref{sec:exp-guarantee} compares the two announcements on the benchmark.

\begin{proof}
(i) Fix a threshold $t > 0$ and a frontier prefix $f$ of the first pass with $p(f) > t$.
At each level below $f$ the prefixes are disjoint with total probability $p(f)$ (A2), so at most $\floor{p(f)/t}$ of
	them exceed $t$, and at most $n_f$ exist.
A prefix that is ever open under tolerance $\delta$ has probability above $(1+\delta)\, s_K(r) \ge (1+\delta)\, s_K(0)$
	and is either a first-pass frontier prefix or a descendant of one.
Hence, at most $\widehat N((1+\delta)\, s_K(0))$ prefixes, \cref{eq:nhat}, are ever open.
If this number is below $M$, the prefixes above the tolerance always fit in a round, they are the heaviest open
	prefixes, so the round expands all of them, and they descend one level per round as in the proof of
	\cref{prop:anydraft}.
After $L-1$ rounds none is open, so $U_R \le (1+\delta)\, s_K(R)$ and \cref{prop:gap} gives the factor $1 + \delta$.
This is the announcement $\delta^*(M)$.

(ii) Take the largest plan for every frontier prefix, pure level-by-level descent.
Since $c_0(f) = 1$, every skeleton node has an item below it, so $c_j(f) \le n_f$, and at most $L-1-\depth(f)$ levels
	lie below $f$, the plan's cost is at most $\min\bigl(\sigma(f),\ 1 + (L-1-\depth(f)) \min(n_f, k_f)\bigr)$, which is
	at most the term of $f$ in \cref{eq:nhat}.
Summing, $S \le \widehat N(t)$, so $\widehat N(t) < M$ gives $\ceil{S/M} + m \le 1 + (L-2) = L-1$: the condition of the
	count without the trie implies ours, hence $\delta^*(M, L) \le \delta^*(M)$.
\end{proof}

\paragraph{Monotonicity of the announcement.}
\begin{proposition}[monotonicity]
\label{prop:plan-more}
\begin{enumerate}[label=\textup{(\roman*)}]
\item $\delta^*(M, Y)$ is non-increasing in $M$ and in $Y$.
\item $\delta^*(M, 2)$ is the smallest $\delta$ for which the skeletons of the prefixes open under $\delta$ fit in one
	pass of $M$ nodes, which is the two-pass bound of the skeleton round (\cref{sec:trie}) under a memory budget.
\end{enumerate}
\end{proposition}

Part (i) says that more memory or more passes never make the guarantee worse.
Part (ii) covers the case $Y = 2$, in which one fallback round remains, so only the plan $\ell = 0$ fits and the pool is
	the union of the skeletons of the open prefixes.

\begin{proof}
(i) The bound $\lceil S/M \rceil + m$ of \cref{prop:rounds} is non-increasing in $M$, and the condition $B \le Y - 1$
	relaxes as $Y$ grows, so the smallest feasible $\delta$ cannot increase in either.

(ii) With $Y = 2$ one round remains, so $m = 0$ and $\lceil S/M \rceil \le 1$ with $S_0(f, t) = \sigma(f)$.
Hence, the skeletons of the prefixes above $t$ must fit in $M$, and the smallest such $\delta$ is the announcement.
\end{proof}

%% file: sections/appendix_C_computing_announcement.tex
\section{Computing the announced factor}
\label{app:compute}

\Cref{eq:deltaplan} defines the announced factor as the smallest $\delta$ whose round bound fits the budget.
This appendix explains what in that computation depends on the request, what can be stored ahead of time, and how the
	minimum is found.

\paragraph{Precomputed and per-request quantities.}
The round bound $R^*(t)$ of \cref{eq:rstar} is built from two kinds of quantities.
The counts $c_j(f)$ and the skeleton sizes at each level below a prefix $f$ depend only on the catalog trie.
They are computed once for every internal node, with the skeleton sizes of each level sorted in decreasing order and
	stored as prefix sums, so that the sum of the $k$ largest is one array lookup.
The frontier $F_0$, its probabilities $p(f)$ and the score $s_K(0)$ depend on the request and are known after the first
	pass.
The threshold $t$ enters $R^*(t)$ in only two places: it decides which prefixes are open, $p(f) > t$, and it sets the
	integer $k_f = \floor{p(f)/t}$ of each open prefix.
For a fixed $f$, the cost $S_\ell(f, t)$ of \cref{eq:plan} is therefore a function of the integer $k_f$ alone.
Moreover, this function stops changing once $k_f$ is large enough.
Once $k_f$ reaches $K_f = \max_j c_j(f)$, every term $\min(c_j(f), k_f)$ equals $c_j(f)$ and the sum of the $k_f$
	largest skeleton sizes is the sum of the whole level, so a larger $k_f$ changes nothing.
One evaluation of $R^*(t)$ is thus a gather of one row per open prefix from the stored tables, followed by a minimum
	over the caps $m \le L-2$, and it costs $O(|F_0| \cdot L)$ operations.
If desired, the table $T_f[m][k] = \min_{\ell \le m} S_\ell(f, k)$ for $k \le K_f$ can also be stored per node, which
	reduces the evaluation to lookups; its total size is at most $(L-1) \sum_f \sigma(f)$ entries.

\paragraph{Bisection.}
$R^*(t)$ is non-increasing in $t$, so the thresholds that fit the budget form an interval that extends to the right.
The implementation bisects on $t$ between $s_K(0)$ and $\max_{f \in F_0} p(f)$, keeps the feasible upper end of the
	bracket, and stops after a fixed number of iterations, twenty in our runs.
The returned threshold is always feasible, so the announcement it yields is valid, and it lies within $2^{-20}$ of the
	bracket above the exact infimum.

\paragraph{Exact search over the breakpoints.}
Because $t$ acts only through the open set and the integers $k_f$, the function $R^*(t)$ is a step function.
It can change only at the thresholds where some $k_f$ changes or some prefix closes, namely at the \emph{breakpoints}
\beq{
B = \bigl\{\, p(f)/k : f \in F_0,\ 1 \le k \le K_f \,\bigr\} \cup \bigl\{\, p(f) : f \in F_0 \,\bigr\},
\label{eq:breakpoints}
}
and it is constant between consecutive breakpoints, since $k_f(t) = k$ for $t \in (p(f)/(k+1),\, p(f)/k]$.
Breakpoints with $k > K_f$ are omitted because the cost has saturated there.
Hence, $|B| \le \sum_{f \in F_0} (K_f + 1)$, which is a few thousand on the benchmark.
After sorting $B$, a binary search finds the first breakpoint whose $R^*$ fits $Y - 1$ in $\lceil \log_2 |B| \rceil$
	evaluations, and the result is exact.
One detail follows from the strict inequality in the definition of an open prefix: the thresholds that fit form an
	interval that is open at its lower end, so $\delta^*$ in \cref{eq:deltaplan} is an infimum, and the planned run uses
	the first representable threshold above the breakpoint, which changes the factor by a rounding unit.

\paragraph{All budgets in one sweep.}
The breakpoints also give the announced factor for every budget in a single sweep.
Walking $B$ in increasing order, each step either lowers one $k_f$ by one or closes one prefix, so each $S^*_m(t)$ is
	updated by a single table difference per cap $m$, and $R^*(t)$ is recomputed as a minimum over $m$.
One sweep therefore yields $R^*(t)$ at every breakpoint in $O(|B| \cdot L)$ time after the sort.
From this curve one reads $\delta^*(M, Y)$ for every $Y$, the number $Y_{\min}(M)$, and, since only the ceiling in
	\cref{eq:rstar} depends on $M$, the same quantities for several memory budgets without recomputation.
The same sweep serves the re-planning after each round, applied to the frontier $F_r$.

\paragraph{Computational cost.}
On the benchmark, the frontier has hundreds of prefixes and $B$ a few thousand entries, so either method takes
	microseconds to milliseconds on a CPU, against a model pass of milliseconds on a GPU.
The choice between bisection and the exact search is therefore a matter of exactness, not of latency.

%% file: sections/appendix_D_details_of_experiments.tex
\section{Details of experiments}
\label{app:exp}

\paragraph{Data, SIDs and model.}
We use Amazon Reviews 2014, category Beauty, with the 5-core filter and the leave-one-out split, which gives 22,363
	users, 12,101 items and 198,502 interactions.
Item text (title, brand, category, description) is embedded with a sentence encoder, and an RQ-VAE with three codebooks
	of 256 codes each, trained with $\beta = 0.25$ and $k$-means initialization, assigns the first three codes, and a
	collision code separates items with equal content codes.
The catalog trie has 256, 9,043 and 11,792 internal nodes at depths 1 to 3.
Of these, 78\% of the depth-2 and 98\% of the depth-3 nodes have a single child, so the skeleton of the catalog has
	2,533 nodes.
The generator is a four-layer decoder-only transformer ($d = 128$, four heads, learned positions, about a million
	parameters) trained with the next-token loss on flattened SID histories of up to 20 items.
Its Hit@10 under exact decoding is 0.055, which is in the range of published results for this setup.

\paragraph{Evaluation protocol.}
For every test user, one tree pass over all 21,091 internal nodes of the catalog trie gives the model's exact
	conditional distribution at every node.
Every decoder, draft and metric in the paper is then a deterministic replay on these tables, which is what makes it
	possible to check every proven statement on every request.
We also verified the replay against the real model.
Tree passes over ordinary sequences agree to within $10^{-5}$ in the largest logit difference, and \cref{alg:main} on
	the real model returns \ModelK on every test user with the number of passes that the replay predicts.

\paragraph{Drafts.}
The model-free drafts are built from the leave-one-out training parts of the sequences and queried with the history
	stale by one interaction.
Content $k$NN scores every item by its largest cosine similarity, in the L2-normalized content embedding, to any of the
	last five history items.
Co-occurrence counts, for every ordered pair of items at most five positions apart in a training sequence, how often the
	second follows the first, row-normalizes the counts, and scores an item by the recency-weighted sum of these rates
	over the last five history items.
Matrix factorization is implicit-feedback weighted ALS with 64 factors, confidence weight 40 and 15 sweeps on the binary
	user--item matrix, with the user's learned vector scored against every item.
All of them exclude the history items themselves and keep the top $N$.
The quantized user sequence is the top-$N$ of a copy of the model whose two-dimensional weights are rounded, per output
	row, to FP4 (E2M1), run on the same one-interaction-old history; its weights are 11\% off in relative Frobenius norm
	and its top-10 agrees with the full model's on 69\% of requests.
An FP8 (E4M3) copy, with weights 2.6\% off and 92\% top-10 agreement, gave a sequence indistinguishable from the full
	model's in every measure, within 0.01 passes at every size, and is omitted from the figures.
Write a user's sequence as $s_1, \dots, s_{n-1}, s_n$ with $s_n$ the held-out target, so that the request's history,
	which the model decodes from, is $s_1 \dots s_{n-1}$.
The user sequence is the model's top-$N$ for $s_1 \dots s_{n-2}$, which is what a batch job that ran before the newest
	interaction $s_{n-1}$ would have stored, so it is stale by one interaction.
The per-user list cannot contain $s_{n-1}$ without a model call for this user after that interaction, which is the
	online cost the method avoids.
Hence, one interaction is the least staleness any per-user batch artifact has.
Finally, no draft contains information about the held-out item $s_n$.
The flatter models apply a temperature to every conditional distribution of the trained model, and the stored sequence
	of a flatter model is recomputed from that model.

\paragraph{Budgets.}
Under a budget of $M$ nodes per pass the first pass feeds the skeleton members of the candidate trie, added in draft
	order while they fit.
Since most candidates lie on chains below a level-1 cluster, a candidate costs a token only when its cluster or a
	branching prefix on its path is new.
As a result, $M = 16$ holds about 90 candidates of the user sequence, $M = 64$ about 210 and $M = 256$ about 440 of its
	500.
The unplanned run in the ablations is \cref{alg:main} with $M$ nodes per pass, i.e., \bmethod{} with the largest plan
	for every prefix and without line~\ref{ln:spec} of \cref{alg:budget}, which amounts to level-by-level descent with
	no schedule and no speculation.
Every figure of \cref{sec:exp} uses the first 10,000 users.

\begin{figure*}[t]
\centering
\includegraphics[width=0.9\textwidth]{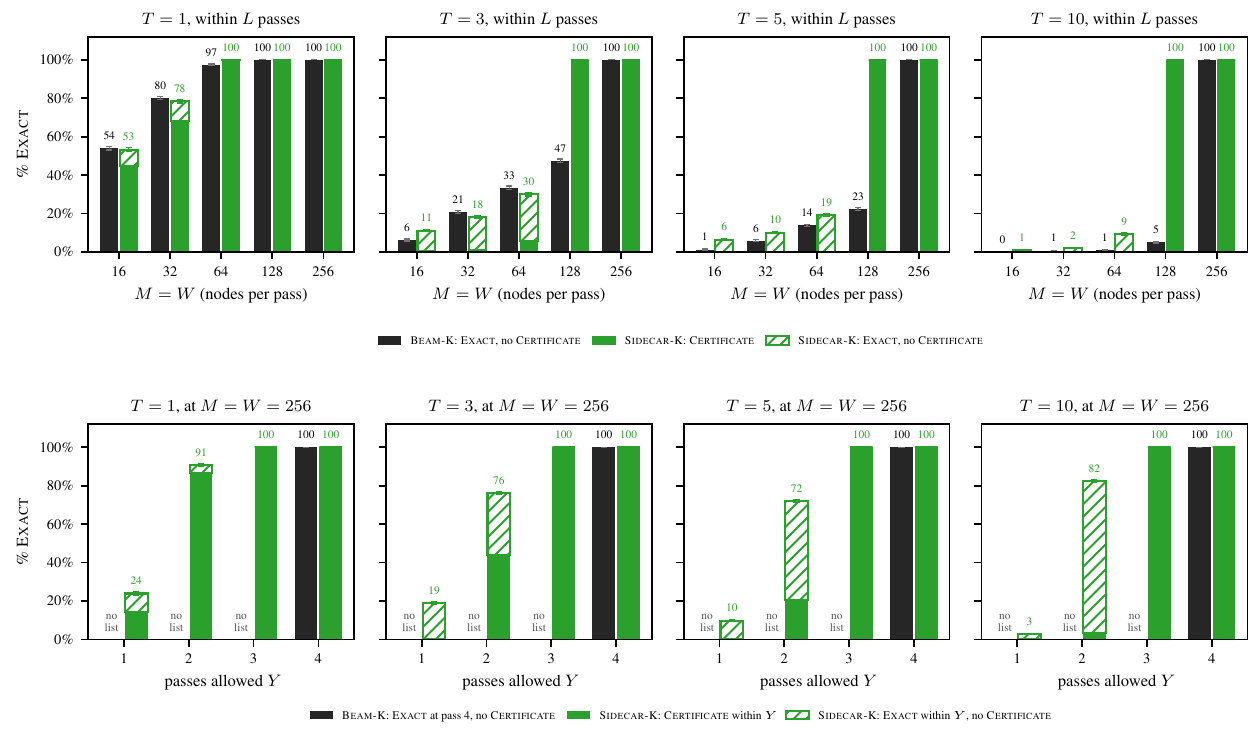}
\caption{\textbf{\Cref{fig:equal} for the MF draft.}
The MF draft is certified after three passes at beam search's memory, but at small memories it is exact somewhat less
	often than \BeamK, on the trained model and at $T = 3$.
A draft that ignores the model therefore still reaches a certified list within beam search's budget, but it does not
	share the advantage that flatness gives the user sequence.}
\label{fig:equal-mf}
\end{figure*}

\paragraph{Equal resources with the MF draft.}
\Cref{fig:equal-mf} repeats the equal-resource comparison of RQ5 (\cref{fig:equal}) for the MF draft.
On the trained model, the MF draft is exact a few percent less often than \BeamK at the smallest memories, because the
	clusters that its first candidates point to are a worse start than the best level-1 codes of the model.
From a moderate memory on, it is exact and certified on nearly every request, like the user sequence.
On the flat models, however, the MF draft does not share the benefit that flatness gives the user sequence
	(\cref{sec:exp-equal}).
The small-subtree items that a flat model ranks highest are not the items that users with the same history went on to
	buy, so the MF draft rarely contains them, and the fallback has to find them.

%% file: sections/appendix_E_faq.tex
\section{Frequently asked questions}
\label{app:faq}

\begin{enumerate}

\item \textbf{Does the draft have to contain the top-$K$?}
No. \Cref{sec:onepass} shows that the first pass is certified when the probability that the draft leaves uncovered is
	below the probability of the $K$-th best item, and its sharp form, \cref{prop:alignment-more}(i), says that the
	certificate holds exactly when the heaviest uncovered prefix is below it.
A draft therefore only has to land in the right prefixes, since the first pass also scores the siblings of the
	candidates without additional tokens, and the fallback finds the remaining items.
This is also why recall against a held-out item is the wrong measure of a draft.
Recall counts whether one item is present, whereas the cost depends on how much probability the draft leaves uncovered.

\item \textbf{What if the draft has fewer than $K$ items, the size of the list?}
Constrained beam search, in TIGER and in the industrial systems of \cref{sec:related} alike, needs a width of at least
	$K$ to return $K$ items, and in practice it runs a width well above $K$.
A draft of at least $K$ items, which a first-stage retriever or a batch job produces without any model call at request
	time, is therefore a mild assumption.
Nevertheless, nothing in the guarantees requires $N \ge K$, and we also test the extreme case $N < K$, in which two
	properties keep the decoder well behaved.
First, the first pass scores more items than the draft holds, since feeding the branching prefixes of one candidate
	scores every item on a chain below them; on our benchmark, a single candidate already fills the list of ten after
	the first pass on nearly every request, except for the user sequence on the flattest models.
Second, on the requests where the first pass nevertheless scores fewer than $K$ items, assumption A5 of
	Appendix~\ref{app:proofs} is violated; the $K$-th best score is then zero, so every frontier prefix is open and the
	fallback first fills the list.
The list therefore remains exact within $L$ passes for every draft (\cref{prop:anydraft}), whereas the announced factor
	is finite only from the round that completes the list on.

\Cref{fig:small-draft} compares the unplanned run with drafts of one to ten items, and of 500 for reference, against
	beam search at equal resources.
On the trained model, the size of the draft barely matters: even a single candidate is about as good as \BeamK at every
	memory and exact far more often at the smallest one.
On the flatter models, however, a draft of a few items gives up most of the advantage that the stored sequence of 500
	entries has at small memory, since a few candidates cannot point to all the clusters that hold the top-10, and at
	intermediate memories it is exact somewhat less often than \BeamK.
From $M = 128$ on, every draft is exact on nearly every request at every temperature, whereas beam search needs a width
	of 256.
Thus, a deployment should keep $N$ well above $K$, but a short draft reduces the share of exact lists only at small
	memory, and it never makes the certificate wrong.

\begin{figure}[t]
\centering
\includegraphics[width=0.95\textwidth]{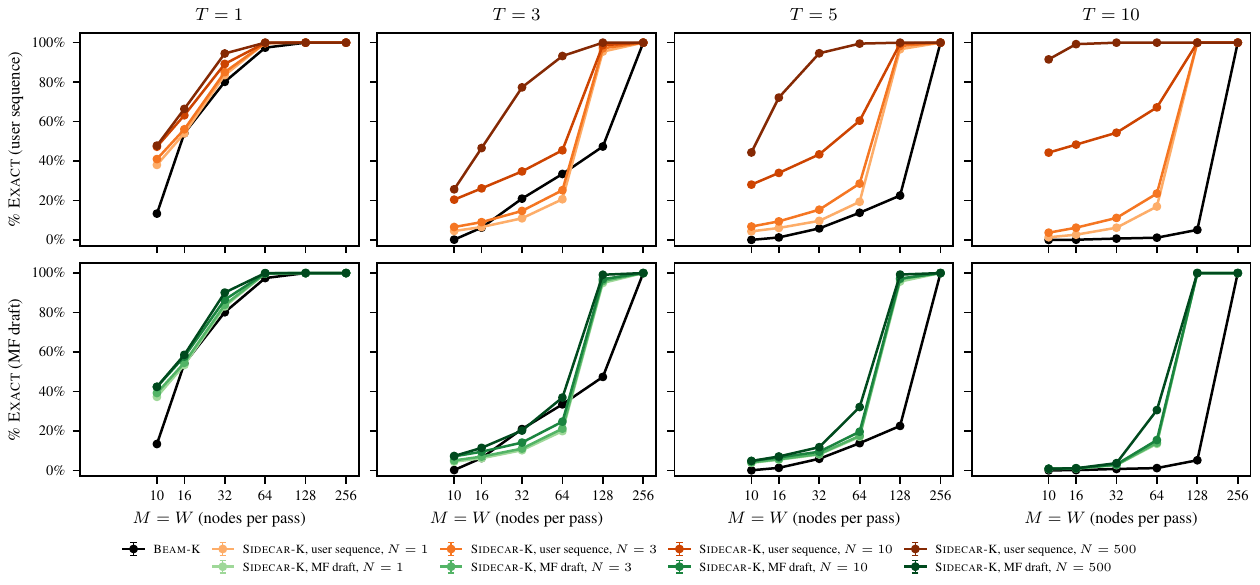}
\caption{\textbf{Drafts smaller than the list}, at beam search's resources ($K = 10$, $Y = L$).
Even a single candidate matches \BeamK on the trained model, and from $M = 128$ on every draft is exact on nearly every
	request, whereas beam search needs the full width.
Below that memory, a short draft gives up most of the advantage that the stored sequence has on the flatter models.}
\label{fig:small-draft}
\end{figure}

\item \textbf{Why does the procedure ever speculate, if speculation costs typical-case accuracy?}
Speculation is what makes a guarantee for the worst case possible.
Without it, the planned run becomes the unplanned run, i.e., \cref{alg:main} with $M$ nodes per pass, and nothing bounds
	its gap before the last pass.
On the instance of \cref{prop:separation}, for example, the unplanned run returns no item of \ModelK, even though the
	budget suffices for exactness.
The reason is that a heavy open prefix whose resolution depth exceeds the remaining rounds cannot be resolved one level
	at a time.
It can only be resolved in time if its skeleton is fed before the model's output at those levels is known.
The schedule of \cref{eq:schedule} forces exactly as much of this as the worst case needs and no more, and the procedure
	fills every other slot with prefixes that are known to be open.
In exchange, the factor $1 + \delta^*$ holds within the budget against the adversary of \cref{sec:budgets}, and the
	measured price in typical-case overlap is small (\cref{sec:exp-price}).

\item \textbf{Is this specific to recommendation?}
The method uses three properties: outputs of fixed length over a prefix-closed vocabulary, constrained scores whose
	prefix probabilities bound the items below, and a draft of plausible outputs.
Any generative retriever over document identifiers has the first two, and most have a first-stage retriever that
	supplies the third.
What is specific to SIDs, then, is the small $L$, which makes the fallback bound of $L$ passes a strong statement, and
	the unary structure of RQ-VAE tries, which makes the skeleton small.

\item \textbf{What would change at production scale?}
The catalog trie would have thousands of level-1 clusters and a skeleton far larger than a pass can hold, so the
	unlimited-memory results no longer apply, and the budgeted results of \cref{sec:budgets} describe the decoder.
The draft then has to be good enough to cover the model's probability, which the stored sequence does by construction,
	and the memory per pass has to cover the open clusters.
The announcement states after the first pass whether it does.

\end{enumerate}